\UseRawInputEncoding
\documentclass{jpsj3}
\usepackage{txfonts}
\usepackage[dvipdfmx]{graphicx}
\usepackage{dcolumn}
\usepackage{bm}
\usepackage{boxedminipage}
\usepackage{amsmath}
\usepackage{graphicx}
\usepackage{amsfonts}
\usepackage{feynmf}
\usepackage{mathrsfs}
\usepackage{braket}
\usepackage{listings}
\usepackage{algorithm}
\usepackage{algorithmic}
\usepackage{color}

\makeatletter
\newcommand\newblock{\hskip .11em\@plus.33em\@minus.07em}
\makeatother

\title{Intrinsic Anomalous Thermal Hall Effect in the Unconventional Superconductor UTe$_2$}

\author{Yukiyasu Moriya, Taiki Matsushita, Masahiko G. Yamada, Takeshi Mizushima \\and Satoshi Fujimoto}
\inst{Department of Materials Engineering Science, Osaka University, Toyonaka, Osaka 560-8531, Japan} 

\abst{{We investigate the intrinsic anomalous thermal Hall effect as an effective probe of the order parameters of non-unitary pairing states of the putative spin-triplet superconductor UTe$_2$. 
We consider all symmetry-allowed non-unitary states under magnetic fields comprehensively, and calculate the anomalous Hall conductivity arising from
time-reversal symmetry-breaking Weyl point nodes of the gap functions, 
assuming three types of Fermi surfaces; ellipsoid-shaped, cylinder-shaped, and ring-shaped ones, which are predicted by previous band calculations.
We find the nonzero anomalous thermal Hall conductivity in the certain ratio of the two irreducible representations. 
The results may be utilized for future exploration of pairing states of UTe$_2$.}
}

\begin{document}
\maketitle

\section{Introduction}
    The recently discovered heavy fermion superconductor UTe$_2$ has attracted much attention in the condensed matter physics.~\cite{UTe2_Butch} UTe$_2$ exhibits some characteristic properties, including an extremely large upper critical field beyond the Pauli limit,~\cite{UTe2_Dai_Aoki,UTe2_Nakamine,UTe2_extreme} field-reentrant superconductivity,~\cite{UTe2_extreme,UTe2_Field-reentrant} small reduction of the NMR Knight shift.~\cite{UTe2_Nakamine} These indicate that the spin-triplet pairng seems to be realized in the superconducting state. In common with other uranium-based ferromagnetic superconductors like UGe$_2$, URhGe, and UCoGe, UTe$_2$ shows strong Ising anisotropy along $a$-axis from the magnetic susceptibility experiment at low temperature.~\cite{UTe2_Butch,UTe2_2006,FMSC_review} On the other hand, the temperature dependence of the magnetization and electrical resistivity in the normal state does not indicate a phase transition to a magnetically ordered state, but rather suggest that 
    UTe$_2$ is situated in the vicinity of the metallic ferromagnetic critical point. Although UTe$_2$ itself is paramagnetic, the behavior of the $H_{c2}$ curve is similar to that of UCoGe and URhGe, suggesting that the ferromagnetic fluctuations mediates forming the Cooper pair.~\cite{UTe2_Butch} In addition, the superconducting transition temperature $T_c$ is about 1.6K, which is rather higher than that of other uranium-based ferromagnetic superconductors. Therefore, UTe$_2$ can be an ideal platform to study pure spin-triplet superconductors. 　
    
    However, the symmetry and the gap structure of UTe$_2$ which determine the topological properties, remain controversial.~\cite{UTe2_Dai_Aoki_2022} Previous studies such as supecific heat, magnetic penetration depth support the existence of point nodes.~\cite{UTe2_Metz,UTe2_Kittaka,UTe2_Shibauchi} In particular, the magnetic field penetration depth measurement indicates low-energy quasiparticle excitations away from the symmetry axes (the $k_b$ and $k_c$ axes), ruling out the possibility of a single-component odd-parity states and suggesting the non-unitary state.~\cite{UTe2_Shibauchi} In addition, the multiple superconducting phases have been observed under pressure and magnetic field, suggesting a multi-component order parameter.~\cite{UTe2_pressure} These results suggest a non-unitary pairing in UTe$_2$, and in this case, Weyl point nodes in the superconducting gap behave as Weyl fermions, resulting the Weyl superconductivity in UTe$_2$.

    In this paper, we show the results of the numerical calculation of the intrinsic anomalous thermal Hall effect (ATHE).~\cite{PhysRevB.61.10267, PhysRevLett.108.026802,ATHE_Sumiyoshi} The ATHE, where a thermal Hall current occurs without an external magnetic field, is a basic property of Weyl superconductors.~\cite{doi:10.7566/JPSJ.85.072001} In particular, the intrinsic effect originates from Berry curvature arising from point nodes in non-unitary states. For the band structure of the normal state, we assume the following three-types of Fermi surfaces. In the first one, an ellipsoid-shaped Fermi surface is assumed as a minimal model that reflects the anisotropy of the crystal structure of UTe$_2$. Note that the possibility of an ellipsoid-shaped Fermi surface surrounding the $\Gamma$ point is discussed in the previous band calculation studies.~\cite{UTe2_Harima, UTe2_ARPES} Of course, we need a more realistic model but the Fermi surface of UTe$_2$ has not yet been observed experimentally. According to the band calculation including the Coulomb interaction $U$, cylinder-shaped and ring-shaped Fermi surfaces have been proposed.~\cite{UTe2_Yanase,UTe2_quasi2D, PhysRevB.100.134502,UTe2_Shishidou} Therefore, the second and third types of Fermi surfaces are assumed to be cylinder-shaped and ring-shaped, respectively. {It is noted that the intrinsic ATHE does not depend on the details of the band structure such as the density of states, because it is determined only by the distribution of the Berry curvature, i.e. the position of monopole charges carried by  Weyl point nodes on the Fermi surfaces at sufficiently low temperatures for which Eqn.(\ref{eq1}) holds. Therefore, it is sufficient to use normal band models which reproduce the qualitative shape of the Fermi surface obtained by band calculations for UTe$_2$ (See Sec.2.2).} Our results directly reflect the topology of each $d$-vector and are useful for the distinction of order parameters based on future thermal conductivity measurements.

    The organization of this paper is as follows. In Sec.2, we introduce the pairing states considered in this study. In Sec.3, we present a theoretical model and parameters we use in our numerical calculation. In Sec.4, we show our results comparing the cases with the magnetic field along $a$-, $b$-, and $c-$axes. We summarize our study in Sec.5. In Appendix, we present the details of the numerical calculation and also, gap node condition of each irrecudible representations (IRs).
\begin{table}[]
    \caption{List of irreducible representations and basis functions for $D_{2h}$.}
    \begin{center}
    \label{t1}
    \begin{tabular}{llc}
    \hline \hline
    IR       & gap structure & basis functions                                 \\ \hline
    $A_u$    & full gap      & $k_a \hat{a}$,\ $k_b \hat{b}$,\ $k_c \hat{c}$   \\
    $B_{1u}$ & point node    & $k_b \hat{a}$ ,\ $k_a \hat{b}$                  \\
    $B_{2u}$ & point node    & $k_a \hat{c}$ ,\ $k_c \hat{a}$                  \\
    $B_{3u}$ & point node    & $k_c \hat{b}$ ,\ $k_b \hat{c}$                  \\ \hline \hline
    \end{tabular}
    \end{center}
\end{table}

\begin{table}[]
    \caption{Possible admixtures of IRs for $D_{2h}$ under magnatic fields.}
    \begin{center}
    \label{t2}
    \begin{tabular}{ccc}
    \hline \hline
    $\bm{H} \parallel a$-axis & $\bm{H} \parallel b$-axis & $\bm{H} \parallel c$-axis    \\ \hline
    $A_u + B_{3u}$            & $A_u + B_{2u}$            & $A_u + B_{1u}$               \\
    $B_{1u} + B_{2u}$         & $B_{1u} + B_{3u}$         & $B_{2u} + B_{3u}$            \\ \hline \hline
    \end{tabular}
    \end{center}
\end{table}

\section{Theoretical model}
    In this section, we describe the model and method used in calculations. 

    \subsection{{Model Hamiltonian}}
         Firstly, the BdG Hailtonian of the system is given by
        \begin{equation}
            H(\bm{k}) = \begin{pmatrix}
                \epsilon(\bm{k}) \hat{\sigma_0} & \hat{\Delta}(\bm{k}) \\
                -\hat{\Delta}^*(-\bm{k}) & -\epsilon(\bm{k}) \hat{\sigma_0} \\
                \end{pmatrix}.
            \label{eqs1}
        \end{equation}
        Here, $\sigma_0$ is the identity matrix, $\epsilon(\bm{k})$ is the energy band in the normal state, and $\hat{\Delta}(\bm{k})$ is a gap function. Particularly, in the case of spin-triplet superconductors, by using $d$-vector({see Sec.2.3 in more detail}), $\hat{\Delta}(\bm{k})$ is represented as,
        \begin{equation}
            \hat{\Delta}(\bm{k}) = \begin{pmatrix}
                -d_a(\bm{k}) + id_b(\bm{k}) & d_c(\bm{k}) \\
                d_c(\bm{k}) & d_a(\bm{k}) + id_b(\bm{k}) \\
                \end{pmatrix}
            \label{eqs2}.
        \end{equation}

    \subsection{{Energy band in the normal state}}
    We consider three models for the energy band $\epsilon(\bm{k})$ which realize three types of the Fermi surfaces, i.e. ellipsoid-shaped, cylinder-shaped, and ring-shaped Fermi surfaces. 
    
        In the case of the ellipsoid-shaped Fermi surface, we used the tight-binding approximation model only including the nearest neighboring(NN) hopping given by
        \begin{equation}
            \epsilon(\bm{k}) = -2(t_a \mathrm{cos}(k_a a) + t_b \mathrm{cos}(k_b b) + t_c \mathrm{cos}(k_c c)) - \mu
            \label{eqs3}
        \end{equation}
        where $t_\mu \ (\mu = a,b,c)$ is the NN hopping energy for $\mu$ direction. We take the parameters as $(t_a,t_b,t_c,\mu)=(1.0,0.80,0.40,-3.5)$.
        
        In the case of cylinder-shaped Fermi surface, we consider the following tight-binding model given by
        \begin{equation}
            \epsilon(\bm{k}) = -2(t_a \mathrm{cos}(k_a a) + t_b \mathrm{cos}(k_b b) + t_c \mathrm{cos}(k_c c)) + \sqrt{4(t_a \mathrm{cos}(k_a a) - t_b \mathrm{cos}(k_b b))^2 + 4V^2} - \mu,
            \label{eqs4}
        \end{equation}
        \begin{equation}
            \epsilon(\bm{k}) = -2(t_a \mathrm{cos}(k_a a) + t_b \mathrm{cos}(k_b b) + t_c \mathrm{cos}(k_c c)) - \sqrt{4(t_a \mathrm{cos}(k_a a) - t_b \mathrm{cos}(k_b b))^2 + 4V^2} - \mu.
            \label{eqs4-2}
        \end{equation}
        Here, Eq.(\ref{eqs4}) and Eq.(\ref{eqs4-2}) corresponds to the electron Fermi surface and hole Fermi surface respectively.
        According to GGA+U, and DFT+U calculations,~\cite{UTe2_Yanase,UTe2_quasi2D} an electron-like and hole-like cylinder-shaped Fermi surfaces are predicted for certain parameters. Eq.(\ref{eqs4}) and (\ref{eqs4-2}) qualitatively reproduce these Fermi surfaces. We take the parameters as $(t_a,t_b,t_c,\mu,V)=(1.0,-0.80,0.10,0.0,0.5)$ in Eq.(\ref{eqs4}) and Eq.(\ref{eqs4-2}).

\begin{figure}[b]
            \includegraphics[width=15cm]{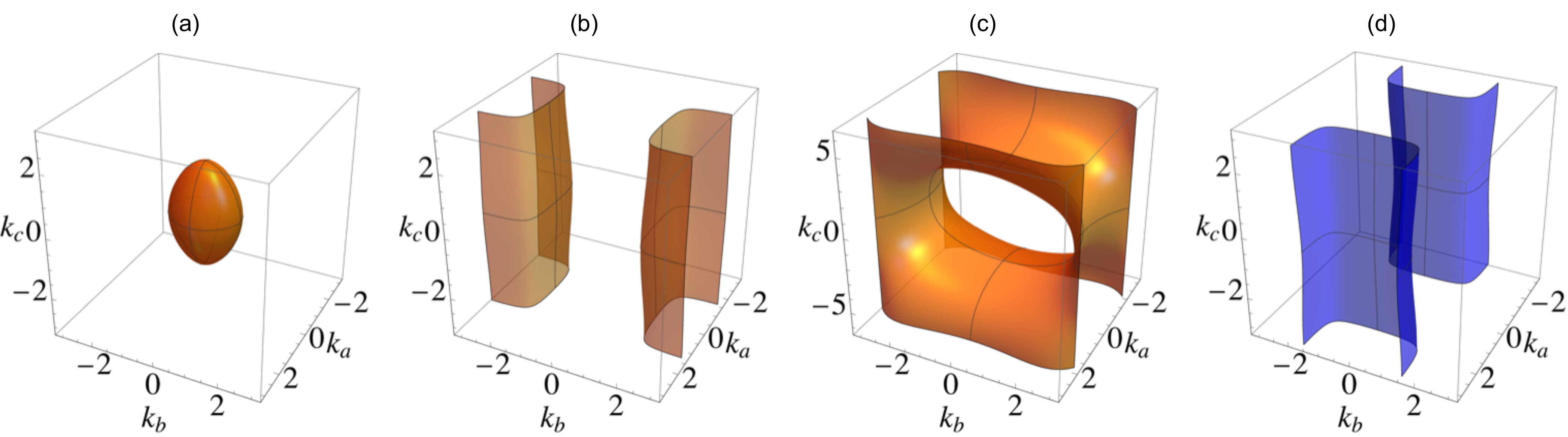}
            \centering
            \caption{Fermi surface used in our calculations. (a)Ellipsoid-shaped Fermi surface given by Eq.(\ref{eqs3}). (b)Cylinder-shaped Fermi surface given by Eq.(\ref{eqs4}). (c)Ring-shaped Fermi surface given by Eq.(\ref{eqs5}).(d)Cylinder-shaped "hole" Fermi surface given by Eq.(\ref{eqs4-2})}
            \label{fermi_surfaces}
        \end{figure}

        In the case of ring-shaped Fermi surface, we used the tight-binding approximation model,
        \begin{equation}
            \epsilon(\bm{k}) = 2t_a \mathrm{cos}(k_a a) + 8 t_U \mathrm{cos}(k_a a/2)\mathrm{cos}(k_b b/2)\mathrm{cos}(k_c c/2) + 2 t_b \mathrm{cos}(k_b b) - \mu + E_0
            \label{eqs5}
        \end{equation}
        which includes the first and second NN hoppings. Here, $t_U$ is the hopping energy of the NN uranium atoms and $t_a, t_b$ are the second NN hopping energy, and $E_0$ is the constant energy shift. The parameters are given by $(t_a,t_U,t_b,\mu,{E_0})=(-1.0,0.78,0.56,{0.0,0.67})$. Fig.\ref{fermi_surfaces} shows the Fermi surfaces for each cases. For simplicity, we set $a=b=c=1$, and the crystal anisotropy of UTe$_2$ is considered only in terms of hopping energy.

        {In this paper, we only consider the non-unitary pairing state. In this case, one of the two Fermi surfaces, which constitute the Kramers pair, has Weyl point nodes and the other does not, i.e. the full gap state. The latter one does not contribute the intrinsic ATHE. Since the Zeemen effect gives rise to a constant shift in the Fermi surface with the Weyl point nodes, it does not affect the qualitative results. Therefore, we do not consider the Zeeman term in the normal band in our calculations. Also, the effect of the Lorentz force-induced thermal Hall effect due to the magnetic flux in the superconductor whose origin is different from intrinsic ATHE is not taken into account here.}

    \subsection{Non-unitary pairing states}
        The point group of UTe$_2$ belongs to $D_{2h}$, and the IRs and the basis functions of the $d$-vector for which odd-parity superconductivity is possible are shown in Table \ref{t1}. Symmetry breaking under magnetic field in each direction can also be considered, and the resulting admixtures of the different IRs are shown in Table \ref{t2}. It is suggested that the results of the magnetic field penetration length measurement for $\bm{H} \parallel a$ are consistent with the non-unitary $A_u + iB_{3u}$ state.~\cite{UTe2_Shibauchi}

        In this paper, we consider six admixtures of IRs for non-unitary pairing states as follows: the $A_u + iB_{3u}$ state and the $B_{1u} + iB_{2u}$ state corresponding to the magnetic field along $a$-axis, the $B_{1u} + iB_{3u}$ state and the $B_{2u} + iA_u$ state corresponding to the magnetic field along $b$-axis, and the $B_{1u} + iA_u$ state and the $B_{2u} + iB_{3u}$ state corresponding to the magnetic field along $c$-axis. The expression for each $\bm{d}$-vector is given by
        \begin{equation}
            \label{eq3}
            \bm{d}_{A_u + iB_{3u}} = (\alpha_1 k_a, \alpha_2 k_b + \beta_1 k_c, \alpha_3 k_c + \beta_2 k_b),
        \end{equation}
        \begin{equation}
            \label{eq4}
            \bm{d}_{B_{1u} + iB_{2u}} = (\gamma_1 k_b + \delta_1 k_c, \gamma_2 k_a, \delta_2 k_a),
        \end{equation}
        \begin{equation}
            \label{eq5}
            \bm{d}_{B_{1u} + iB_{3u}} = (\gamma_1 k_b, \gamma_2 k_a + \beta_1 k_c, \beta_2 k_b),
        \end{equation}
        \begin{equation}
            \label{eq6}
            \bm{d}_{B_{2u} + iA_u} = (\alpha_1 k_a + \delta_1 k_c, \alpha_2 k_b, \alpha_3 k_c + \delta_2 k_a),
        \end{equation}
        \begin{equation}
            \label{eq7}
            \bm{d}_{B_{1u} + iA_u} = (\alpha_1 k_a + \gamma_1 k_b, \alpha_2 k_b + \gamma_2 k_a, \alpha_3 k_c),
        \end{equation}
        \begin{equation}
            \label{eq8}
            \bm{d}_{B_{2u} + iB_{3u}} = (\delta_1 k_c, \beta_1 k_c, \beta_2 k_b + \delta_2 k_a),
        \end{equation}
        where the coefficients $\alpha, \beta, \gamma, \delta$ correspond to $A_u, B_{3u}, B_{1u}, B_{2u}$, respectively. In the following, we focus on the behavior of the thermal Hall conductivity which is determined by the positions of the point nodes. For this purpose, we calculated the thermal Hall conductivity as a function of the ratio $r$ of the amplitudes of the two IRs.

        Here, we will discuss the details of the $d$-vector given by Eq(\ref{eq3})-(\ref{eq8}) in our calculation. First, in the case of the magnetic field along $a$-axis, for the the $A_u + iB_{3u}$ state, we set $(\alpha_1,\alpha_2,\alpha_3,\beta_1,\beta_2) = (r, 1.1r,2.0r,i,i)$. Here, $r = |\alpha_1/\beta_1|$ and $i$ represents the imaginary unit. The parameters for $\alpha_j$ $(j = 1,2,3)$ are set to be consistent with Ising anisotropy with easy-axis parallel to the a-axis. For the $B_{1u} + iB_{2u}$ state, we set $(\gamma_1,\gamma_2,\delta_1,\delta_2) = (r, r, -i, -i)$ and $r = |\gamma_1/\delta_1|$. Next, in the case of the magnetic field along $b$-axis, for the $B_{1u} + iB_{3u}$ state, we set $(\gamma_1,\gamma_2,\beta_1,\beta_2) = (r, r, i, i)$ and $r = |\gamma_1/\beta_1|$. For the $B_{2u} + iA_u$ state, we set $(\alpha_1,\alpha_2,\alpha_3,\delta_1,\delta_2) = (0, i, i, r, r)$ and $r = |\delta_1/\alpha_2|$. We set $\alpha_1 = 0$ to take the Ising anisotropy into account. 
        Finally, in the case of the magnetic field along $c$-axis, for the $B_{1u} + iA_{u}$ state, we set $(\alpha_1,\alpha_2, \alpha_3, \gamma_1,\gamma_2) = (0, i, i, r, r)$ and $r = |\gamma_1/\alpha_2|$. For the $B_{2u} + iB_{3u}$ state, we set $(\beta_1,\beta_2,\delta_1,\delta_2) = (i, i, r, r)$ and $r = |\delta_1/\beta_1|$. In the numerical calculations, the basis of the $d$-vector is replaced as $k_a a \rightarrow \sin{k_a a}$ to be consistent with the tight-binding  model of the normal state. We note that the essential physics remains the same.

    \subsection{The intrinsic anomalous thermal Hall effect}
        Three-dimensional intrinsic anomalous thermal Hall conductivity is given by
        \begin{equation}
            \label{eq1}
            \kappa_{\mu \nu} = \frac{\pi T}{12} \frac{1}{2\pi}\int \mathrm{d}k_\xi C(k_{\xi}).
        \end{equation}
        Here, the subscripts $\mu,\nu,\xi$ correspond to the crystallographic axes $a,b,c$ and $C(k_{\xi})$ represents the momentum-resolved Chern number defined at momentum $k_{\xi}$ which given by
        \begin{equation}
            \label{eq2}
            C(k_{\xi}) = \frac{1}{2 \pi i} \int \mathrm{d}k_{\mu} \mathrm{d}k_{\nu} \mathrm{Tr} \left( P \left(\frac{\partial P}{\partial k_{\mu}} \frac{\partial P}{\partial k_{\nu}} - \frac{\partial P}{\partial k_{\nu}} \frac{\partial P}{\partial k_{\mu}} \right)\right),
        \end{equation}
        where $P$ is the projection operators into the ground-state manifold of occupied bands.~\cite{Kitaev2006} 
        Although the Chern number is also expressed in terms of the Berry curvature in the momentum space, Eq.(\ref{eq2}) is more advantageous for numerical calculations. Since the Weyl points in the gap structure can be a source or drain of the Berry curvature, the intrinsic anomalous thermal Hall conductivity directly reflects detail of the order parameter of Weyl superconductors.


\section{Results of numerical calculations of Intrinsic ATHE}

   In this section, we, first, present the results for a simple ellipsoid-shaped Fermi surface 
        to explain the condition for nonzero $\kappa_{\mu\nu}$ which is deeply related to the positions of Weyl point nodes on the Fermi surface. Then, we present the results for the cylinder-shaped, ring-shaped Fermi surfaces predicted by the band calculations.~\cite{UTe2_Yanase, UTe2_quasi2D, UTe2_Harima}

    \subsection{Ellipsoid-shaped fermi surface}
    {First, we discuss the condition for the non-vanishing intrinsic ATHE which is deeply related to the positions of the Weyl point nodes in the case of the simple ellipsoid-shaped Fermi surface.}

        \subsubsection{{The $A_u + iB_{3u}$ state and the $B_{1u} + iB_{2u}$ state, candidates of pairings in the magnetic field parallel to $a$-axis}}
            \begin{figure}[b]
                \includegraphics[width=9cm]{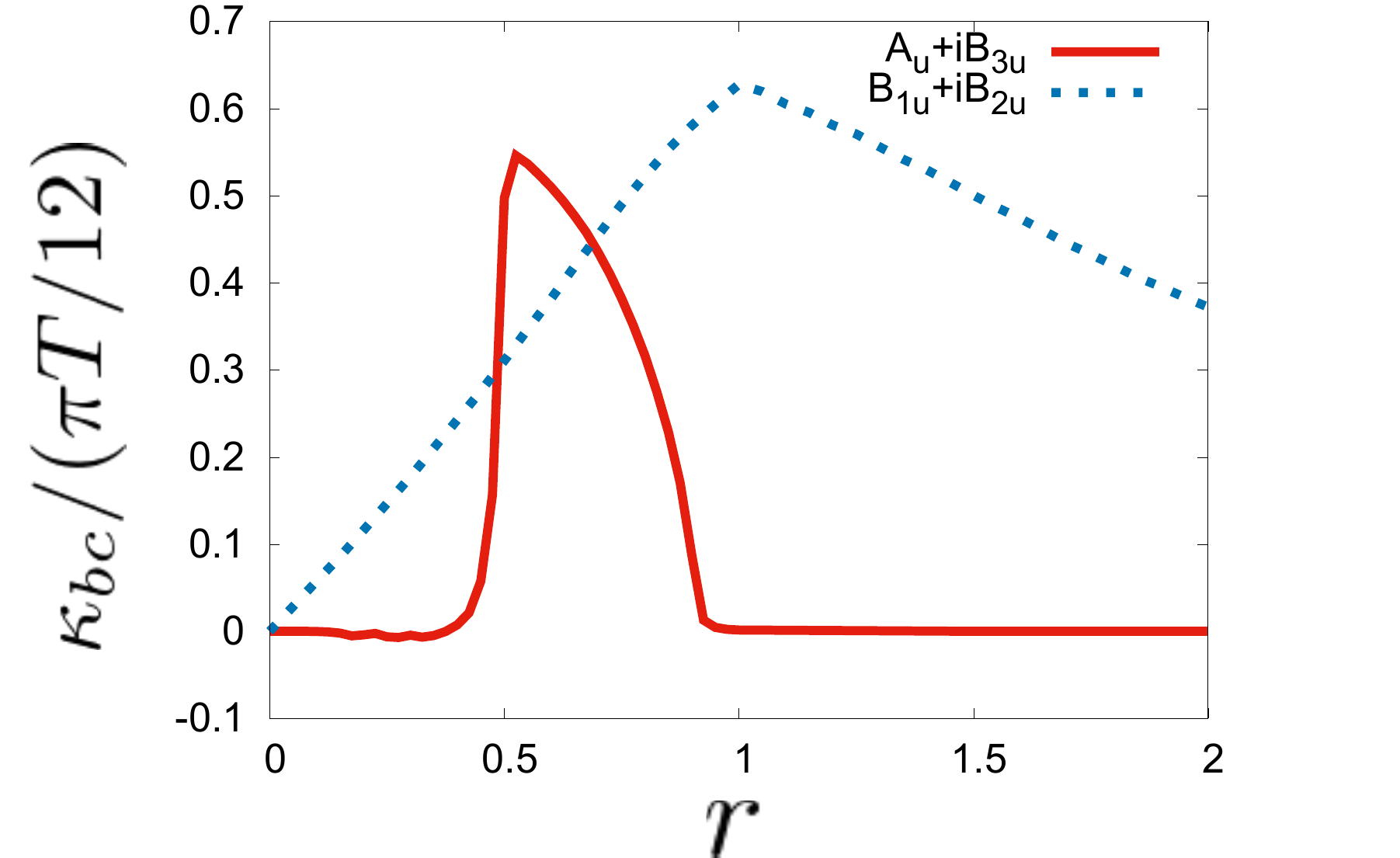}
                \centering
                \caption{The comparison of $r$ dependence of the anomalous thermal Hall conductivity for the $A_u + iB_{3u}$ state(red {solid} line) and the $B_{1u} + iB_{2u}$ state(blue {dotted} line) in the case of ellipsoid-shaped Fermi surface. The $d$-vector and the ratio of two IRs are given by $\bm{d}_{A_u + iB_{3u}} = (\alpha_1 k_a, \alpha_2 k_b + \beta_1 k_c, \alpha_3 k_c + \beta_2 k_b)$ and $r = |\alpha_1/\beta_1|$ for the $A_u + iB_{3u}$ state, $\bm{d}_{B_{1u} + iB_{2u}} = (\gamma_1 k_b + \delta_1 k_c, \gamma_2 k_a, \delta_2 k_a)$ and $r = |\gamma_1/\delta_1|$ for the $B_{1u} + iB_{2u}$ state.}
                \label{comparison_Ha_ellipsoid}
            \end{figure}

            \begin{figure}[b]
                \includegraphics[width=15cm]{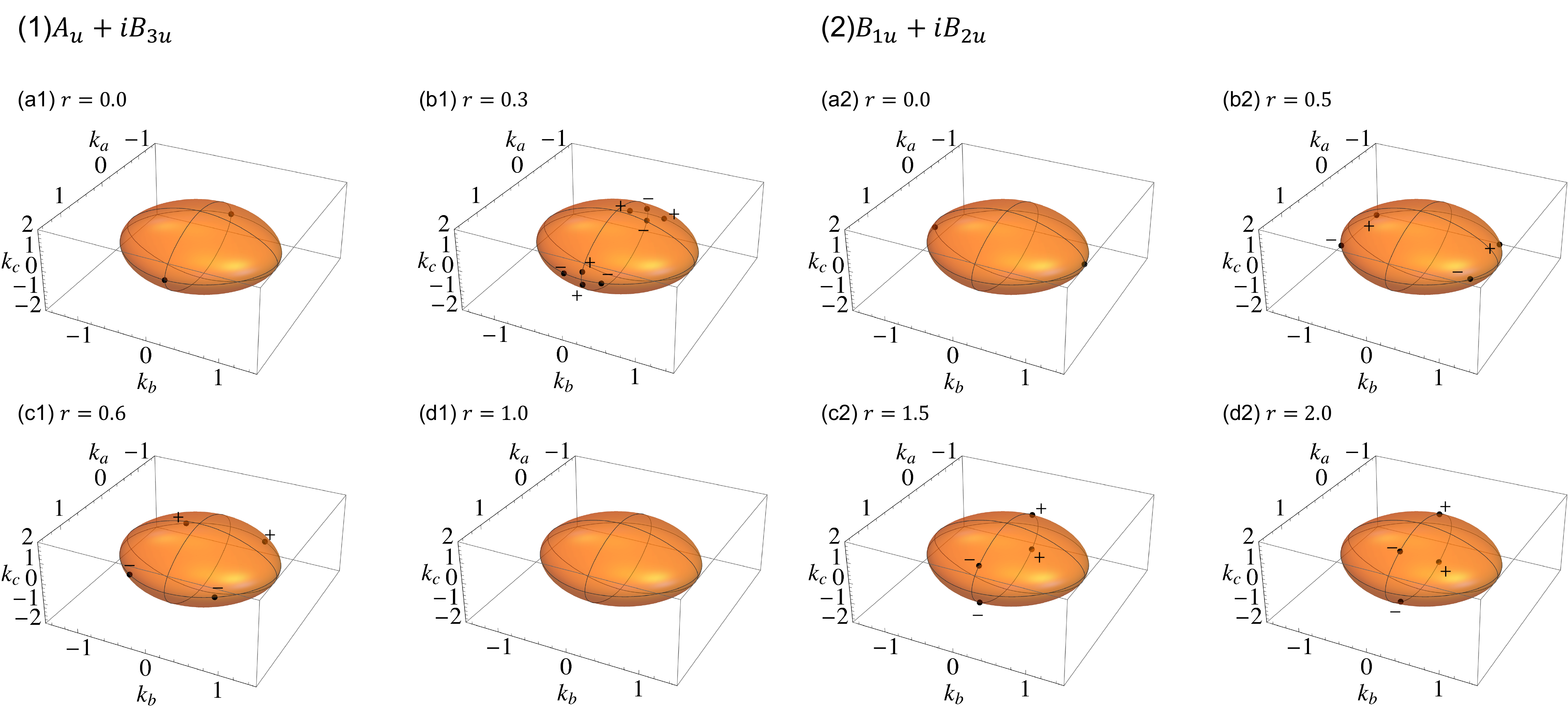}
                \centering
                \caption{The positions of the point nodes for (1)the $A_u + iB_{3u}$ state and (2)$B_{1u}+iB_{2u}$ on the ellipsoid-shaped Fermi surface for several $r$ values. The sign $+(-)$ represents the source(drain) of the Berry curvature.}
                \label{node_comparison_Ha_ellipsoid}
            \end{figure}
            First, the result of the thermal Hall conductivity for the ellipsoid-shaped Fermi surface are shown in Fig.\ref{comparison_Ha_ellipsoid}. The $x$-axis refers to the ratio of the two component of each IRs, i.e. $r = |\alpha_1/\beta_1|$ for $A_u + iB_{3u}$ and $r = |\gamma_1/\delta_1|$ for $B_{1u} + iB_{2u}$ indicating that the larger the value of $r$, the more the $A_u$(or $B_{1u}$) component becomes dominant. In Fig.\ref{comparison_Ha_ellipsoid}, we see that the thermal Hall conductivity has a peak at $r = 0.5$. In Fig.\ref{node_comparison_Ha_ellipsoid}(1), we show the positions of the point nodes on the Fermi surface for the certain values of $r$. The sign $+(-)$ represents the source(drain) of the Berry curvature. 
            
            {In the case of the $A_u + iB_{3u}$ state,} when $r = 0$, i.e. only the $B_{3u}$ state exists, there are two point nodes on the $k_a$-axis. They are Dirac points which do not break the time-reversal symmetry, and thus, the Berry curvature is zero. In the region $0.0 < r \lesssim 0.5$, there are 8 point nodes which break the time-reversal symmetry resulting the finite Berry curvature. In this case, however, the integration of the Berry curvature over the 1st Brillouin zone vanishes, i.e. the thermal Hall conductivity is zero. On the other hand, in the region $0.5 \lesssim r \lesssim 0.9$, among these 8 point nodes, the pair of point nodes on the $k_a k_c$-plane which has opposite sign annihilate. As a result, there are 4 point nodes on the $k_b k_c$-plane. In this case, the Berry curvature does not cancel and we get finite thermal Hall conductivity. This pair annihilation of half number of point nodes is due to the orthorhombic crystal system. We can easily understand that thermal Hall conductivity becomes zero after all point nodes disappear at $r \simeq 0.9$.

            On the other hand, the structure of point nodes for the $B_{1u} + iB_{2u}$ state, shown in Fig.\ref{node_comparison_Ha_ellipsoid}(2), dramatically changes around $r = 1$. In the range of $r < 1$, where $B_{2u}$ is dominant, the point nodes move closer to the $k_a$ axis as $r$ increases. In this case, the distance between the point nodes along the $k_a$ direction increases, i.e., the region where the Chern number is non-zero increases, and the thermal Hall conductivity also increases. On the other hand, in the range of $r > 1$, the point nodes move closer to the {$k_c$} axis as $r$ increases. In this case, the distance between the point nodes becomes smaller, and the opposite happens, resulting in a decrease in the thermal Hall conductivity. 
    
            As shown above, there are important differences in the qualitative behaviors of these two states. For the $A_u + iB_{3u}$ state, when $r > 0.9$, the full gap state is realized and the thermal Hall conductivity is completely zero, but for the $B_{1u} + iB_{2u}$ state, it slowly decays to the limit where $B_{1u}$ is dominant. As mentioned earlier, this originates from the change in the point-node structure, and thus, the behavior of the thermal Hall conductivity provides an important information for the determination of the gap function.

        \subsubsection{{The $B_{1u} + iB_{3u}$ state and the $B_{2u} + iA_u$ state, candidates of pairings in the magnetic field parallel to $b$-axis}}

                \begin{figure}[b]
                    \includegraphics[width=9cm]{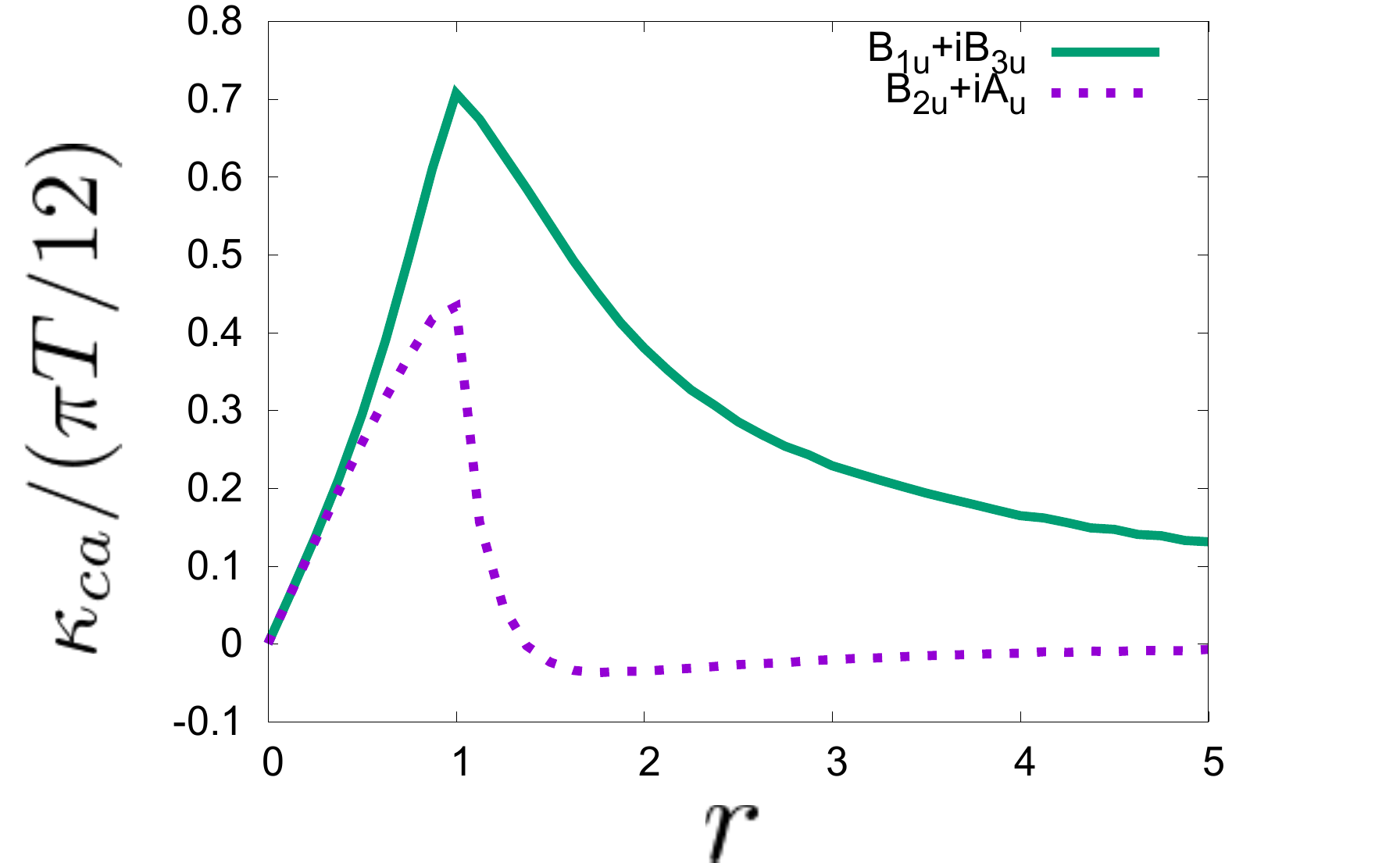}
                    \centering
                    \caption{The comparison of $r$ dependence of the anomalous thermal Hall conductivity for the $B_{1u} + iB_{3u}$ state(green {solid} line) and the $B_{2u} + iA_u$ state(purple {dotted} line) in the case of ellipsoid-shaped Fermi surface. The $d$-vector and the ratio of two IRs are given by $\bm{d}_{B_{1u} + iB_{3u}} = (\gamma_1 k_b, \gamma_2 k_a + \beta_1 k_c, \beta_2 k_b)$ and $r = |\gamma_1/\beta_1|$ for the $B_{1u} + iB_{3u}$ state, $\bm{{d}}_{B_{2u} + iA_u} = (\alpha_1 k_a + \delta_1 k_c, \alpha_2 k_b, \alpha_3 k_c + \delta_2 k_a)$ and $r = |\delta_1/\alpha_2|$ for the $B_{2u} + iA_u$ state.}
                    \label{comparison_Hb_ellipsoid}
                \end{figure}

                \begin{figure}[b]
                    \includegraphics[width=15cm]{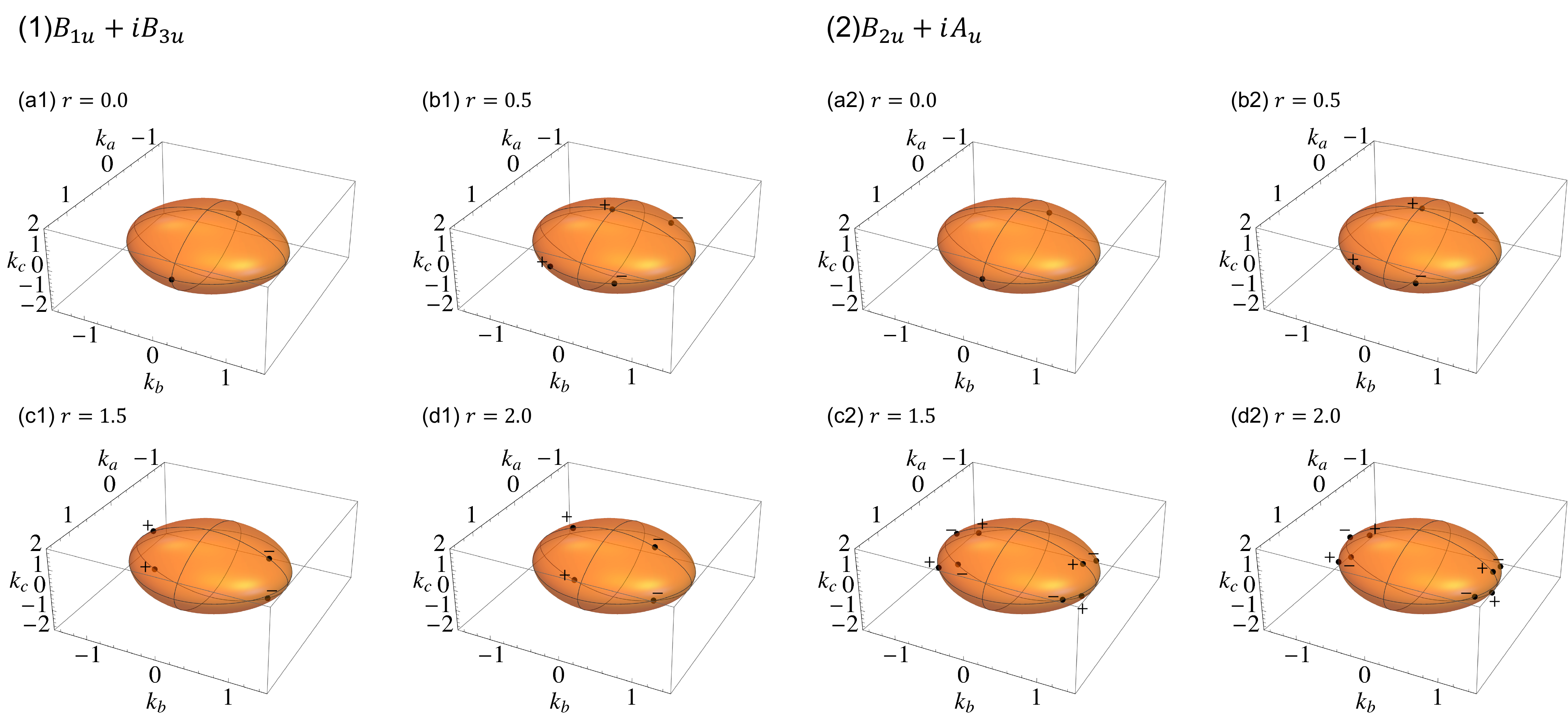}
                    \centering
                    \caption{The positions of the point nodes for (1)the $B_{1u} + iB_{3u}$ state and (2)$B_{2u}+iA_u$ on the ellipsoid-shaped Fermi surface for several $r$ values. The sign $+(-)$ represents the source(drain) of the Berry curvature.}
                    \label{node_comparison_Hb_ellipsoid}
                \end{figure}
            In Fig.\ref{comparison_Hb_ellipsoid}, the horizontal axis represents the ratio of two IRs $r = |\gamma_1/\beta_1|$ for the $B_{1u} + iB_{3u}$ state and $r = |\delta_1/\alpha_2|$ for the $B_{2u} + iA_u$ state, respectively. 
            
            In both cases, thermal Hall conductivity monotonically increases up to $r \simeq 1$, while it shows a quite different behavior for $r \gtrsim 1.0$. The similarities and differences in these behaviors can be explained by the point node structure, as well as former cases. In Fig.\ref{node_comparison_Hb_ellipsoid}, we show the positions of point nodes in each values of $r$. In the region $r < 1.0$, the point nodes split off from the $k_a$ axis in both cases and move closer to the $k_b$ axis as $r$ increases. 
            
            However, the behavior of point nodes changes drastically when $r$ becomes larger than 1.0. In the case of the $B_{2u} + iA_u$ state, the point nodes split from the $k_b$ axis into four, but in the case of the $B_{1u} + iB_{3u}$ state, they split into two. Recalling the case of the $A_u + iB_{3u}$ state, the total Berry curvature cancels out by the integration over the first Brillouin zone when there are eight point nodes and the thermal Hall conductivity is strongly suppressed. On the other hand, in the case of the $B_{1u} + iB_{3u}$ state, the Berry curvature does not cancel, and the thermal Hall conductivity decreases monotonically with increasing $r$ until all nodes disappear on the $k_c$ axis. These results are consistant with our calculations and directly reflect the point-node structure.

        \subsubsection{{The $B_{1u} + iA_u$ state and the $B_{2u} + iB_{3u}$ state, candidates of pairings in the magnetic field parallel to $c$-axis}}
            \begin{figure}[b]
                \includegraphics[width=9cm]{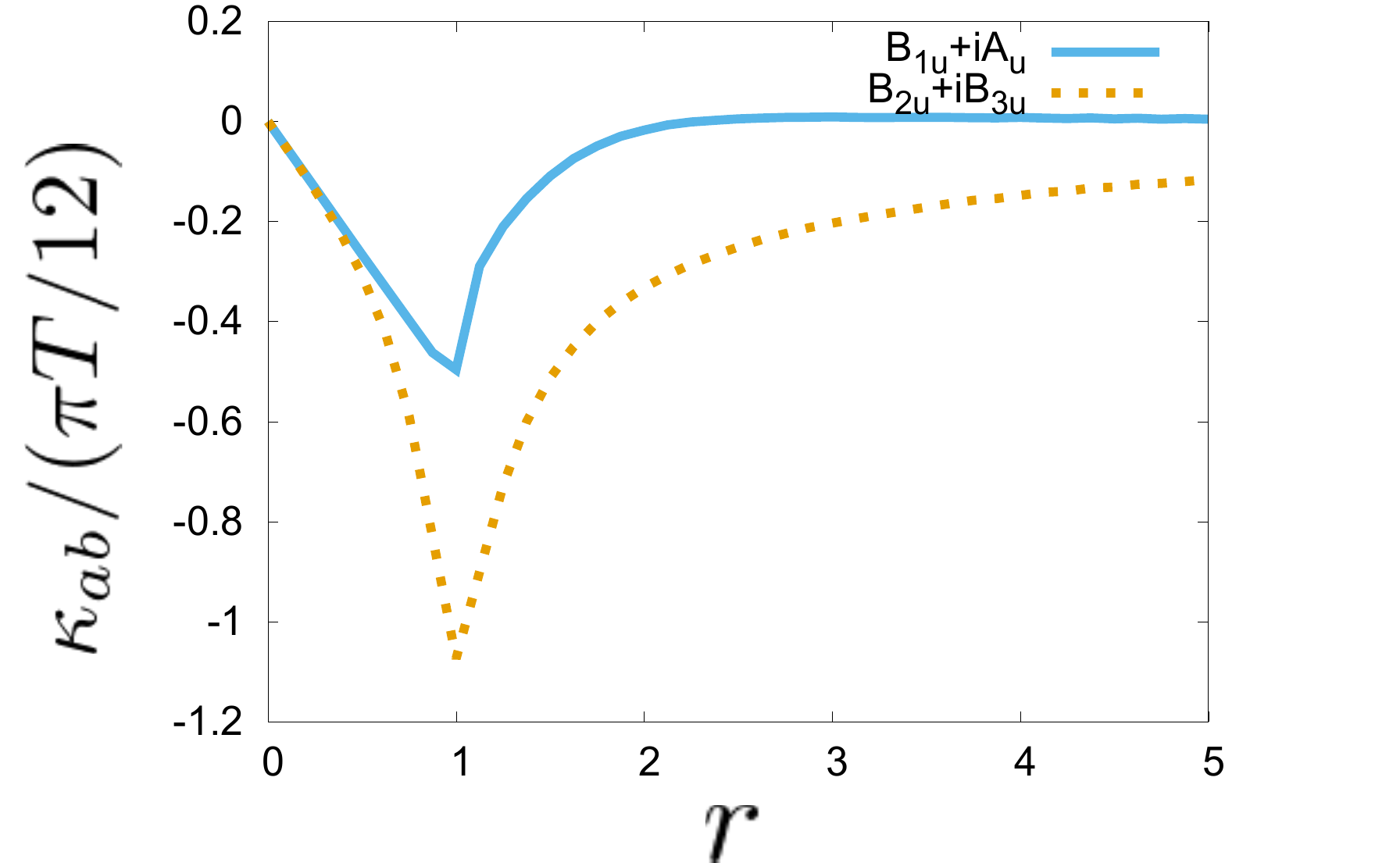}
                \centering
                \caption{The comparison of $r$ dependence of the anomalous thermal Hall conductivity for the $B_{1u} + iA_u$ state(light blue {solid} line) and the $B_{2u} + iB_{3u}$ state(yellow {dotted} line) in the case of ellipsoid-shaped Fermi surface. The $d$-vector and the ratio of two IRs are given by $bm{d}_{B_{1u} + iA_u} = (\alpha_1 k_a + \gamma_1 k_b, \alpha_2 k_b + \gamma_2 k_a, \alpha_3 k_c)$ and $r = |\gamma_1/\alpha_2|$ for the $B_{2u} + iB_{3u}$ state, $\bm{{d}}_{B_{2u} + iB_{3u}} = (\delta_1 k_c, \beta_1 k_c, \beta_2 k_b + \delta_2 k_a)$ and $r = |\delta_1/\beta_1|$ for the $B_{2u} + iB_{3u}$ state.}
                \label{comparison_Hc_ellipsoid}
            \end{figure}
        
                \begin{figure}[b]
                    \includegraphics[width=15cm]{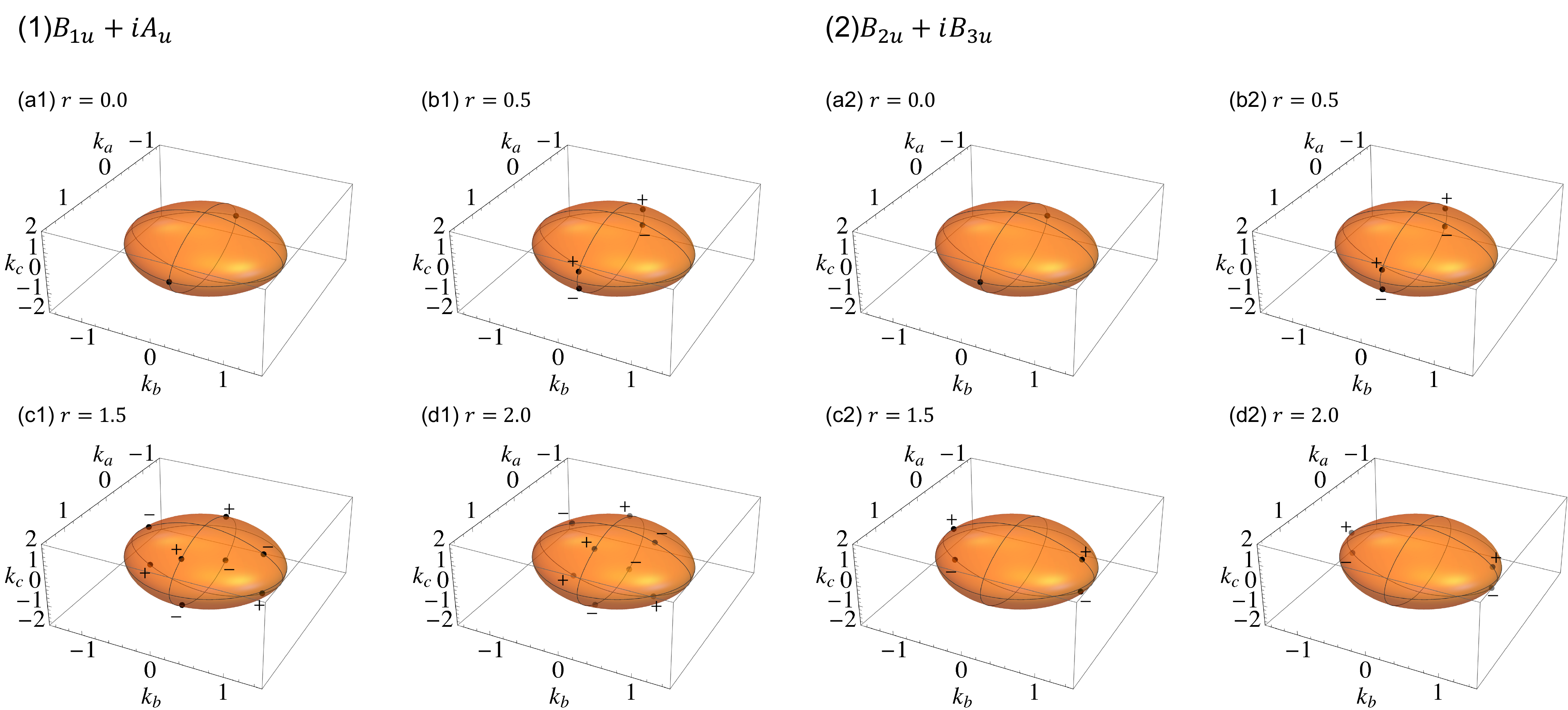}
                    \centering
                    \caption{The position of the point nodes for (1)the $B_{1u} + iA_u$ state and (2)$B_{2u}+iB_{3u}$ on the ellipsoid-shaped Fermi surface for each $r$ values. The sign $+(-)$ represents the source(drain) of the Berry curvature.}
                    \label{node_comparison_Hc_ellipsoid}
                \end{figure}
            In Fig.\ref{comparison_Hc_ellipsoid}, we show the result of the case for an ellipsoid-shaped Fermi surface. In this case, $x$-axis represents the ratio of two IRs $r = |\gamma_1/\alpha_2|$ for the $B_{1u} + iA_u$ state and $r = |\delta_1/\beta_1|$ for the $B_{2u} + iB_{3u}$ state, respectively. 
            
            The overall behavior and the difference between two states are similar to that shown in Fig.\ref{comparison_Hb_ellipsoid}. In other words, as shown in Fig.\ref{node_comparison_Hc_ellipsoid}, the rate of decay of the anomalous thermal Hall conductivity with respect to $r$ is due to the difference in point node structure at $r > 1$.

    \subsection{Electron-like and hole-like cylinder-shaped Fermi surfaces}
        According to the recent band calculations,\cite{UTe2_Yanase,UTe2_quasi2D,UTe2_Harima} the possibility of two cylinder-sheped Fermi surfaces composed of electron and hole bands is discussed. We, here, consider the anomalous thermal Hall conductivity for this type of Fermi surfaces. 
        \subsubsection{{The $A_u + iB_{3u}$ state and the $B_{1u} + iB_{2u}$ state, candidates of pairings in the magnetic field parallel to $a$-axis}}
                \begin{figure}[b]
                    \includegraphics[width=15cm]{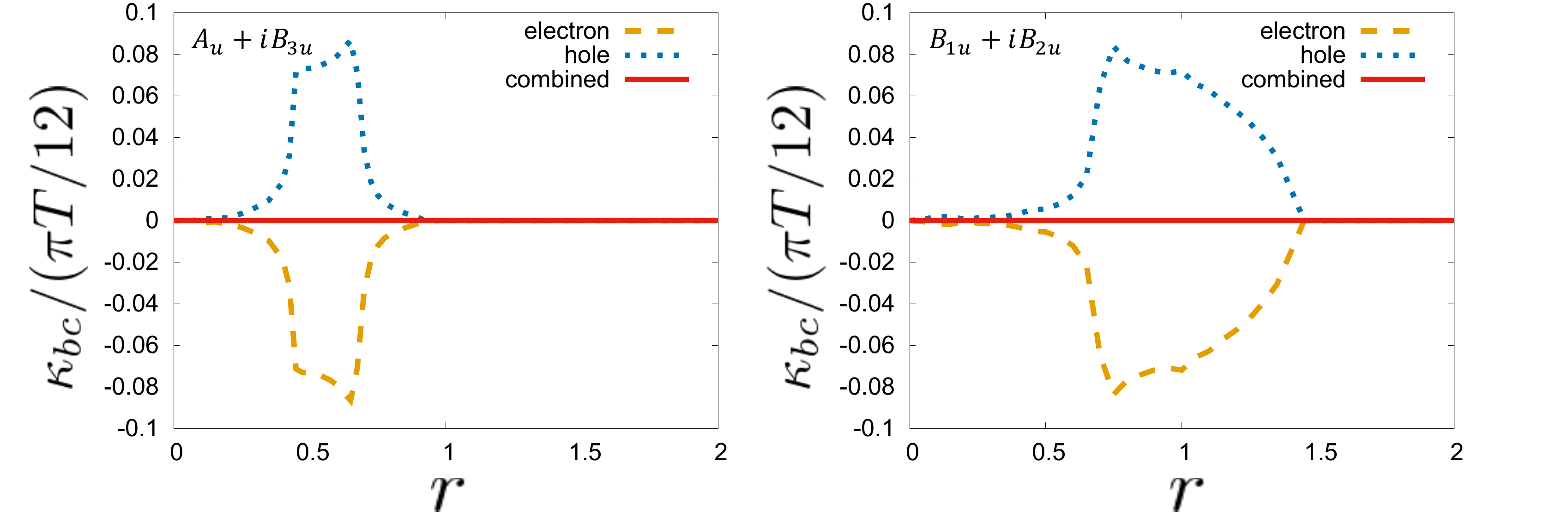}
                    \centering
                    \caption{The comparison of $r$ dependence of the anomalous thermal Hall conductivity for the $A_u + iB_{3u}$ state(left figure)) and the $B_{1u} + iB_{2u}$ state(right figure) in the case of two cylinder-shaped Fermi surfaces.}
                    \label{comparison_Ha_cylinder}
                \end{figure}

                \begin{figure}[b]
                    \includegraphics[width=15cm]{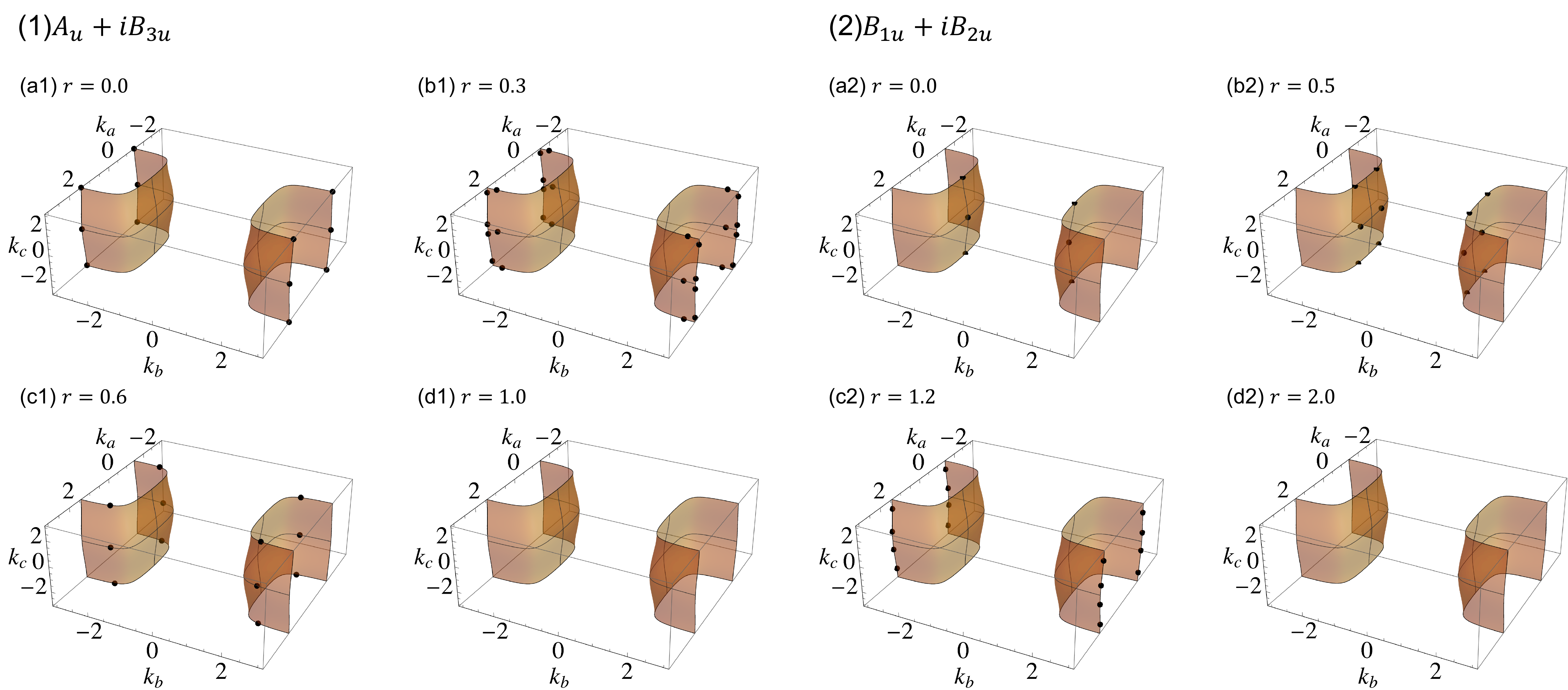}
                    \centering
                    \caption{The positions of the point nodes for (1)the $A_u + iB_{3u}$ state and (2)$B_{1u}+iB_{1u}$ on the cylinder-shaped Fermi surface for several $r$ values. }
                    \label{node_comparison_Ha_cylinder}
                \end{figure}
            The result of the thermal Hall conductivity is shown in Fig.\ref{comparison_Ha_cylinder}. Focusing on the electron surface contributions(blue line), we find that the finite thermal Hall conductivity can be obtained in the region $0.5 \lesssim r \lesssim 0.9$ in the case of the $A_u + iB_{3u}$ state, as in the case of ellipsoid-shaped Femri surface condidered before. The distinct difference from the ellipsoid-shaped case is the behavior in the vicinity of the peak. This is because the shape of the Fermi surface is flat compared to the ellipsoid-shaped case. In particular, the $k_a$-coordinates of the point nodes shown in Fig.\ref{node_comparison_Ha_cylinder}(1) does not change monotonically from $r \simeq 0.5$, when half of the point nodes disappear, to $r \simeq 0.6$. This is reflected in the fact that the peaks in our results are relatively flat in this range. As in the case of the $A_u + iB_{3u}$ state, the behavior of the thermal Hall conductivity for the $B_{1u}+iB_{2u}$ state is not monotonous like the ellipsoid-shaped case, but a peak-like structure is obtained in the range of $r$ where the $B_{1u}$ component becomes dominant. The important result here is that at $r \simeq 1.5$ the point nodes completely disappear, i.e., the full gap state is realized and the thermal Hall conductivity is completely zero. 

            In both cases, however, the overall behavior is completely suppressed by the contribution from the hole surface. The hole surface is given by Eq.(\ref{eqs4-2}), and the qualitative behavior in thermal hall conductivity is explained by the opposite sign of the results of the electron band. The complete cancellation of the ATHE is due to the fact that the system is a compensated metal. In contrast to the normal Hall effect, which depends on the relaxation time and the effective band mass as well as the band curvatures, the intrinsic ATHE is determined solely by the Berry curvature. Because of this reason, the compensated metal band given by Eqs.(\ref{eqs4}) and (\ref{eqs4-2}) result in the total cancellation.

        \subsubsection{{The $B_{1u} + iB_{3u}$ state and the $B_{2u} + iA_u$ state, candidates of pairings in the magnetic field parallel to $b$-axis}}
                \begin{figure}[b]
                    \includegraphics[width=15cm]{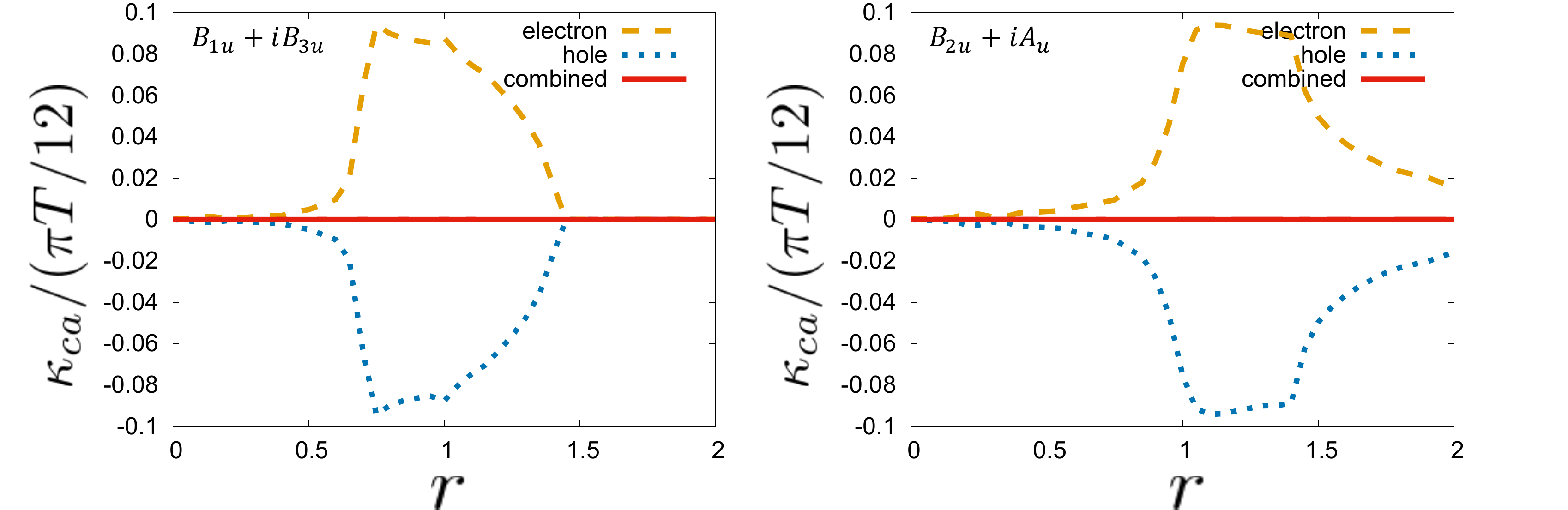}
                    \centering
                    \caption{The comparison of $r$ dependence of the anomalous thermal Hall conductivity for the $B_{1u} + iB_{3u}$ state(left figure) and the $B_{2u} + iA_u$ state(right figure) in the case of two cylinder-shaped Fermi surfaces.}
                    \label{comparison_Hb_cylinder}
                \end{figure}

                \begin{figure}[b]
                    \includegraphics[width=15cm]{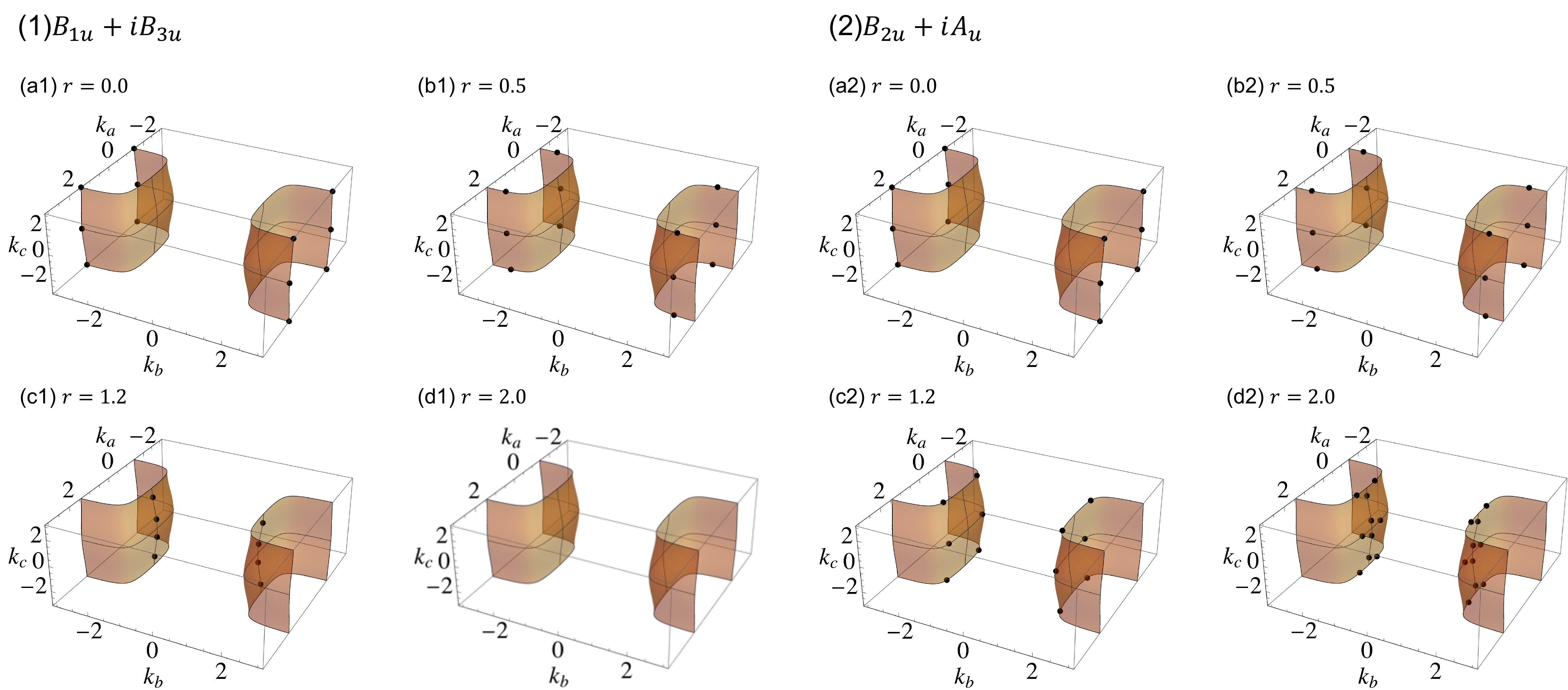}
                    \centering
                    \caption{The position of the point nodes for (1)the $B_{1u} + iB_{3u}$ state and (2)$B_{2u}+iA_u$ on the cylinder-shaped Fermi surface for several $r$ values. }
                    \label{node_comparison_Hb_cylinder}
                \end{figure}
            The result of the case with the cylinder-shaped Fermi surface are shown in Fig.\ref{comparison_Hb_cylinder} and we show the positions of the point nodes for several $r$ values in Fig.\ref{node_comparison_Hb_cylinder}. In this case also the thermal Hall conductivity results in the total cancellation due to the band structure of the compensated metal.

        \subsubsection{{The $B_{1u} + iA_u$ state and the $B_{2u} + iB_{3u}$ state, candidates of pairings in the magnetic field parallel to $c$-axis}}
                \begin{figure}[b]
                    \includegraphics[width=15cm]{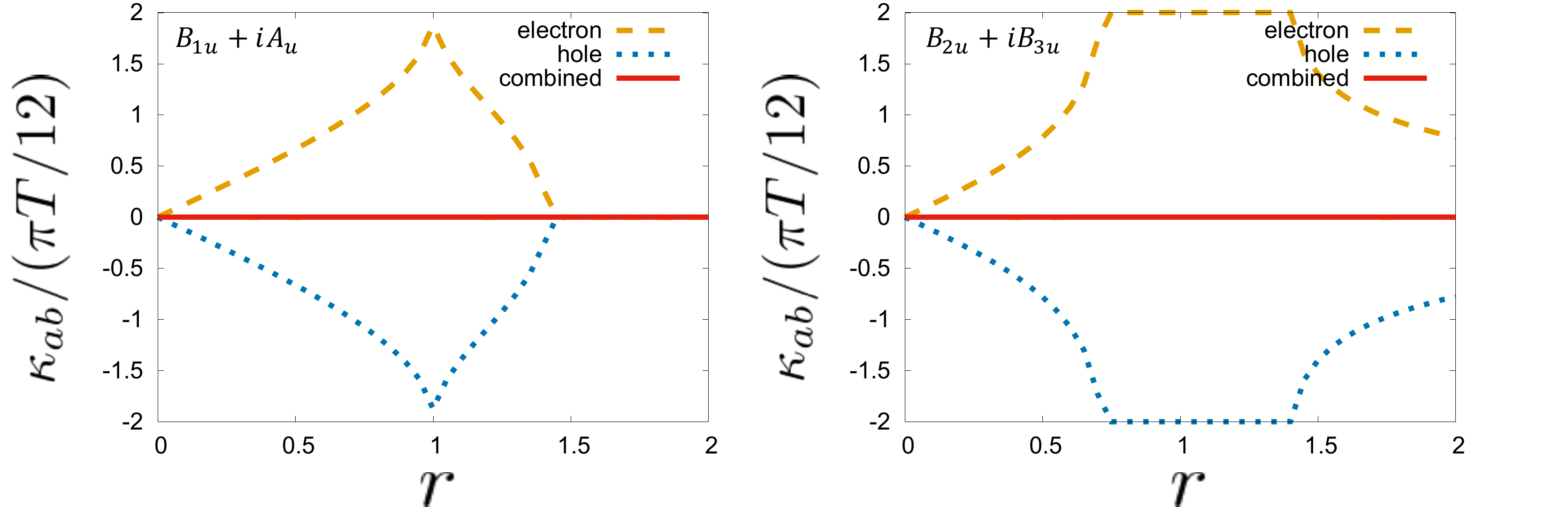}
                    \centering
                    \caption{The comparison of $r$ dependence of the anomalous thermal Hall conductivity for the $B_{1u} + iA_u$ state(left figure) and the $B_{2u} + iB_{3u}$ state(right figure) in the case of two cylinder-shaped Fermi surfaces.}
                    \label{comparison_Hc_cylinder}
                \end{figure}

                \begin{figure}[b]
                    \includegraphics[width=15cm]{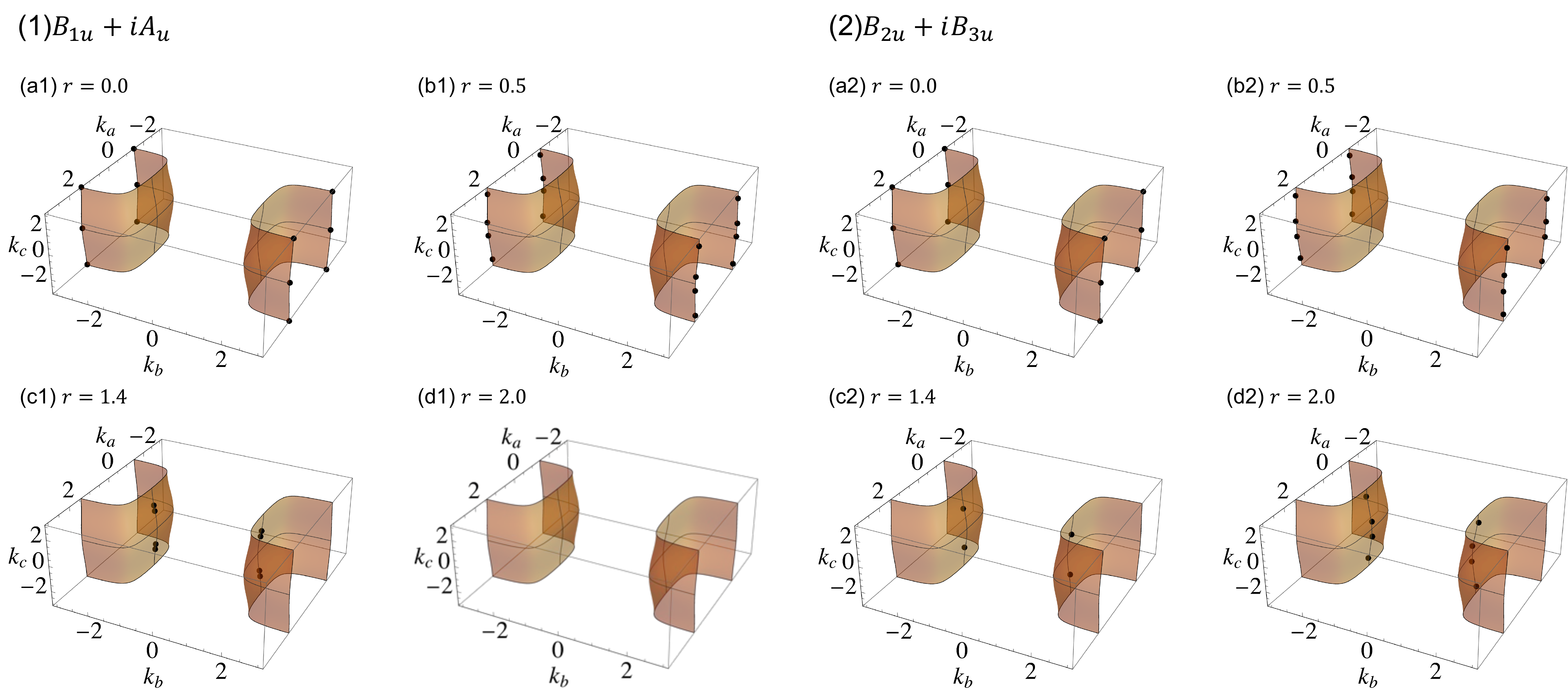}
                    \centering
                    \caption{The position of the point nodes for (1)the $B_{1u} + iA_u$ state and (2)$B_{2u}+iB_{3u}$ on the cylinder-shaped Fermi surface for each $r$ values. }
                    \label{node_comparison_Hc_cylinder}
                \end{figure}
            The results in this case are shown in Fig.\ref{comparison_Hc_cylinder}. As in the cases discussed so far, the contribution from the hole surface completely suppresses the thermal Hall conductivity. 

    \subsection{Ring-shaped Fermi surface + cylinder-shaped hole Fermi surface}
     {   Here, we consider the combined results of a ring-shaped electron-like Fermi surface and a cylinder-shaped hole-like Fermi surface. According to the band calculations\cite{UTe2_Yanase,UTe2_quasi2D}, these two Fermi surface may coexist for certain parameters.   The following results are the combination of the contribution from the cylinder-shaped and ring-shaped contributions which are calculated independently.} 

        \subsubsection{{The $A_u + iB_{3u}$ state and the $B_{1u} + iB_{2u}$ state, candidates of pairings in the magnetic field parallel to $a$-axis}}
            \begin{figure}[b]
                \includegraphics[width=15cm]{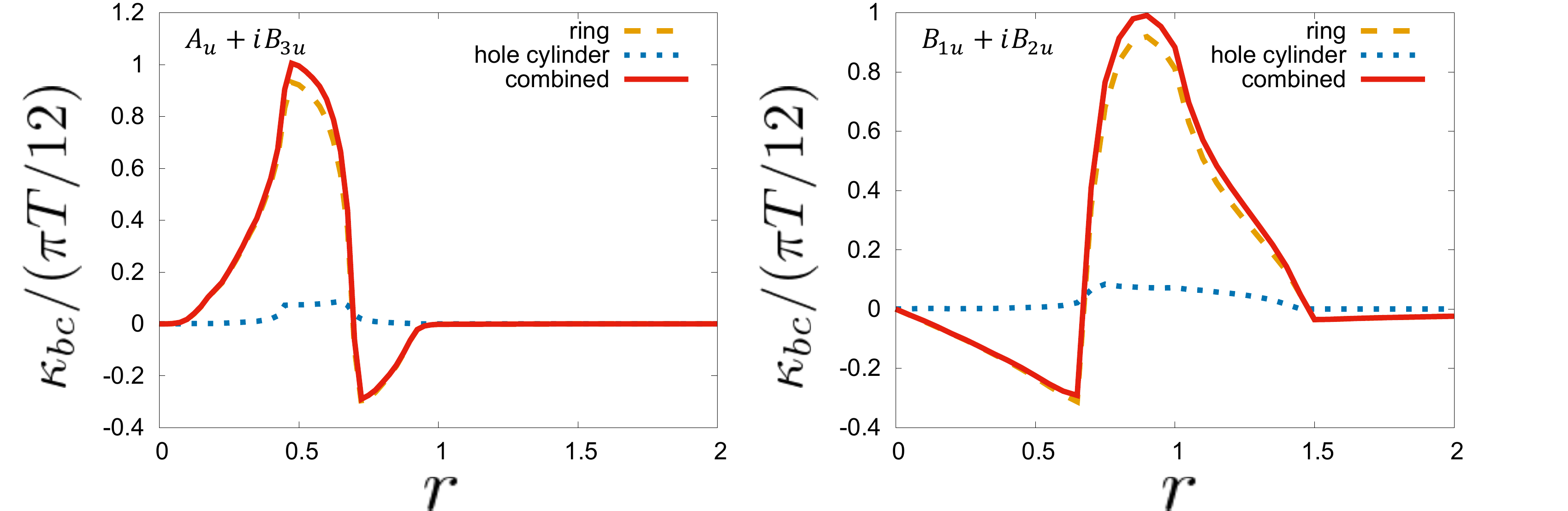}
                \centering
                \caption{The comparison of $r$ dependence of the anomalous thermal Hall conductivity for the $A_u + iB_{3u}$ state(left figure) and the $B_{1u} + iB_{2u}$ state(right figure) in the case of cylinder-shaped Fermi surface + ring-shaped Fermi surface.}
                \label{comparison_Ha_combo}
            \end{figure}
            For the $A_u + iB_{3u}$ state, the peak width is wider than in the ellipsoid-shaped case, but the overall behavior is similar. The distinct difference from the ellipsoid-shaped Fermi surface is the sign change at $0.7 \lesssim r \lesssim 0.9$. In the case of $\kappa_{bc}$, the distance of the point nodes in the $k_a$ direction and whether they are sources or drains of Berry curvature are important. Such a sign change suggests a drastic change of the configuration of the point nodes, i.e., a reversal in the width of the regions with positive and negative Chern number. Note that the number of point nodes in the Brillouin zone is lager than that of the ellipsoid-shaped and cylinder-shaped cases, and the model dependence is also stronger. However, all cases discussed so far suggest that finite thermal Hall conductivity are expected to be obtained in the range of $r$ such that the half of the point nodes which split into four by the admixture of the two IRs are annihilated and the half remain. 
            
            On the other hand, for the $B_{1u} + iB_{2u}$ state, the overall behavior is also similar to the ellipsoid-shaped case and the sign change is also seen. This result also depends on the details of the shape of the Fermi surface, but it is interesting to note that in the ellipsoid-shaped case the thermal Hall conductivity is finite up to the limit where $B_{1u}$ is dominant, whereas in the ring case it is completely zero at $r \simeq 1.5$. This is because the point node structure changes at $r \simeq 1.5$, i.e., the Weyl point node partially disappears. 

            A comparison of the peak intensities between cylinder-shaped Fermi surface and ring-shaped Fermi surface shows that the ring-shaped one gives the dominant contribution. Therefore, if the Fermi surface of UTe$_2$ consists of the ring-shaped and the cylinder-shaped ones, the contribution to the thermal Hall conductivity is dominated by the ring-shaped Fermi surface.

        \subsubsection{{The $B_{1u} + iB_{3u}$ state and the $B_{2u} + iA_u$ state, candidates of pairings in the magnetic field parallel to $b$-axis}}

                \begin{figure}[b]
                    \includegraphics[width=15cm]{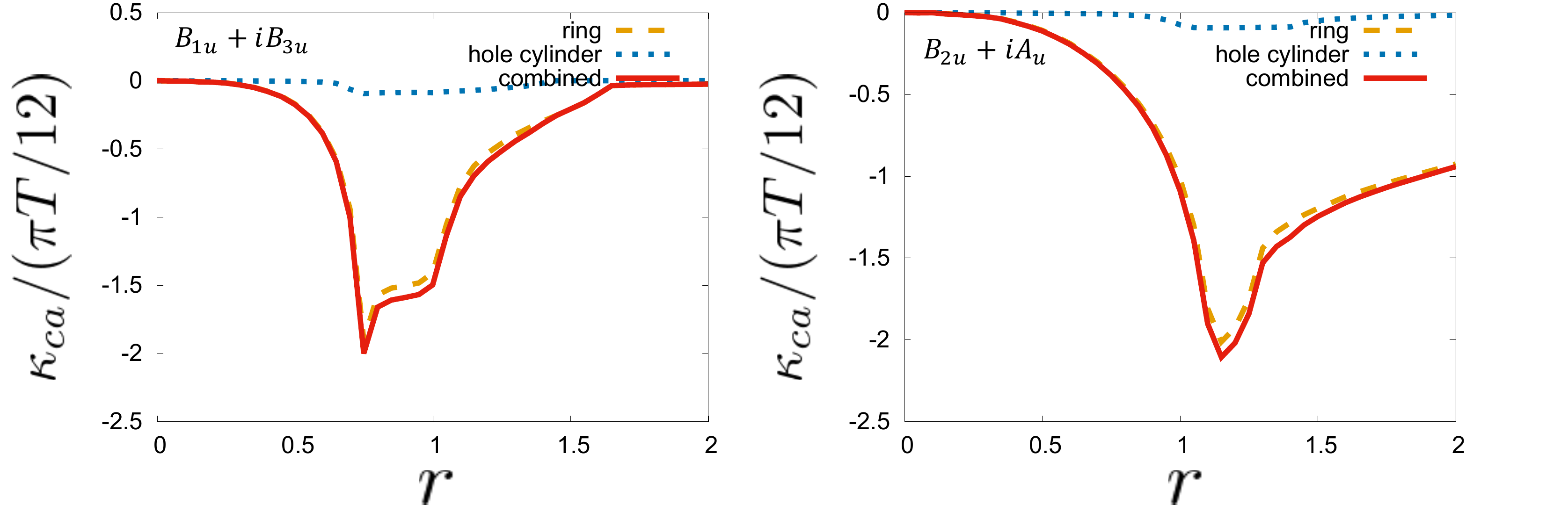}
                    \centering
                    \caption{The comparison of $r$ dependence of the anomalous thermal Hall conductivity for the $B_{1u} + iB_{3u}$ state(left figure) and the $B_{2u} + iA_u$ state(right figure) in the case of cylinder-shaped Fermi surface + ring-shaped Fermi surface.}
                    \label{comparison_Hb_combo}
                \end{figure}
            As in the case under the magnetic field along $a$-axis, the contribution from the ring-shaped Fermi surface is dominant compared to that from the cylinder-shaped Fermi surface. Therefore, we discuss the contribution of the ring-shaped Fermi surface.

            As shown is Fig.\ref{comparison_Hb_combo}, the behavior of the thermal Hall conductivity is the same for $r < 0.5$ because the point-node structure is also the same for the two states, as in the case of the ellipsoid-shaped and cylinder-shaped cases. For $r > 0.5$, the position of the two split point nodes become closer as $r$ increases, however, at a faster rate in the $B_{1u} + iB_{3u}$ case than the $B_{2u} + iA_u$ state(due to the fact that the $k_a$ component in the $A_u$ state is set to zero to take into Ising anisotropy account). This results in different peak positions between the two states. The important difference in this case is the sharp peak at $r \simeq 0.75$ in the the $B_{1u} + iB_{3u}$ state. This is due to the annihilation of the point nodes at this point. The flat behavior in the range $0.8 < r < 1.0$ is comes from the fact that the point nodes are located only inside the ring and the $k_b$ coordinates of them moves slightly with increasing of $r$. On the other hand, in the case of the $B_{2u} + iA_u$ state, when $r > 1$, the point nodes gradually gather inside of the ring. As $r$ increases, the point nodes inside the ring split into two (the $B_{1u} + iB_{3u}$ state) or four (the $B_{2u} + iA_u$ state) as before, which can be confirmed as the difference in the behavior of the thermal Hall conductivity due to the cancellation of Berry curvature.

        \subsubsection{{The $B_{1u} + iA_u$ state and the $B_{2u} + iB_{3u}$ state, candidates of pairings in the magnetic field parallel to $c$-axis}}
                \begin{figure}[b]
                    \includegraphics[width=15cm]{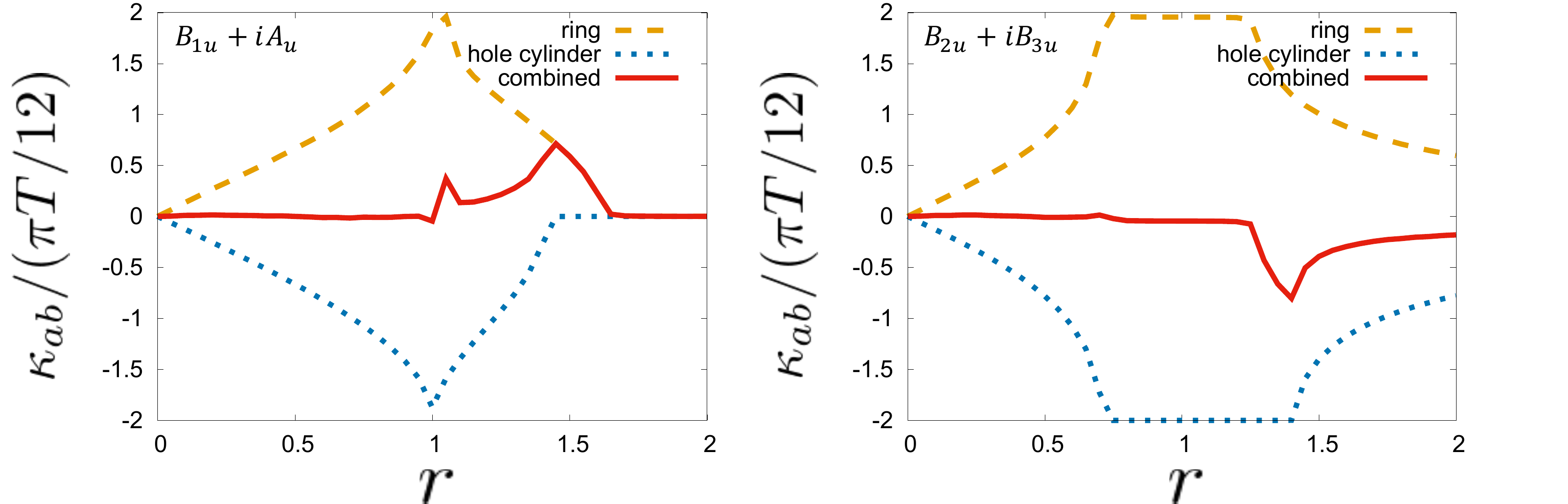}
                    \centering
                    \caption{The comparison of $r$ dependence of the anomalous thermal Hall conductivity for the $B_{1u} + iA_u$ state(left figure) and the $B_{2u} + iB_{3u}$ state(right figure) in the case of cylinder-shaped Fermi surface + ring-shaped Fermi surface.}
                    \label{comparison_Hc_combo}
                \end{figure}
            {The results shown in Fig.\ref{comparison_Hc_combo} are similar to those for the cylinder-shaped case shown in Fig.\ref{comparison_Hc_cylinder}.
            This is because that the contribution from the ring-shaped Fermi surface for $\kappa_{ab}$ is dominated by the partially cylindrical parts surrounding the $k$-points $(0,\pm\frac{4}{3}\pi,0)$. However, in this case, the exact cancelation between electron and hole surfaces does not occur, because of 
            the deviation from the cylindrical shape of the electron surface for large $|k_c|$. }
            
           { In the case of the $B_{1u} + iA_u$ state, thermal Hall conductivity has two peaks around $r \simeq 1$ and $r\simeq 1.4$.
            Also, the vanishing of $\kappa_{ab}$ from the contribution of the ring-shaped Fermi surface for $r > 1.6$ is not due to the full gap state at this point as in the case of the cylinder, but to cancellation of the Berry curvature. }
            
           {On the other hand,  in the $B_{2u} + iB_{3u}$ state,  the anomalous thermal Hall conductivity has only one peak. The difference of the peak structure between the $B_{1u} + iA_u$ and $B_{2u} + iB_{3u}$ state can be utilized for the distinction of these two states. We note that the sign of $\kappa_{ab}$ depends on the relative phase between two gap function which are hybridized via a magnetic field. The relative phase depends on the detail of the microscopic pairing mechanism.  Thus, the difference of the sign $\kappa_{ab}$ between the two states shown in Fig.\ref{comparison_Hc_combo} is model-dependent. } 

            In the case with the magnetic field along the $a$ and $b$-axes discussed so far, the contribution from the ring-shaped Fermi surface is dominant compared to that of the cylinder-shaped Fermi surface.  However, this point is quite different from the case with the magnetic field along the $c$-axis. In this case, the electron contribution of the ring-shapes Fermi surface is mostly cancelled by the hole contribution of the cylinder-shaped one, partially because of the compensated metal character.

\section{Discussion}
    In this study, we performed theoretical calculations focusing on the intrinsic anomalous thermal Hall effect as an effective probe of the order parameter of non-unitary paring states in UTe$_2$. As a result, we found that the thermal Hall conductivity depends on the positions and the total number of time-reversal symmetry-breaking point nodes (Weyl points) in the gap structure. In particular, in the case of non-unitary $d$-vectors  which are composed of two different IRs, the peak structure is seen in a certain region of the ratio of the amplitudes of the two representations, reflecting the fact that there are regions where Berry curvature cancellation does not occur due to the orthorhombic crystal structure. Our results imply that the $r$-dependence of the anomalous thermal Hall conductivity exhibits distinct behaviors for different symmetries of the paring states. It is expected that the mixing ratio of different IRs in the non-unitary states can be controlled by applied magnetic fields. Thus, the systematic investigations of the magnetic field-dependence of the anomalous thermal Hall conductivity provide information on paring states. 
   
    {As explained before, the intrinsic ATHE is completely determined by the positions of monopole charges carried by Weyl point nodes on the Fermi surfaces, and not affected by other details of the band structure such as the density of states.
    Thus, we would like to emphasize that in the case of a compensated metal with electron and hole cylinder-shaped Fermi surfaces, 
 the exact cancellation of the contributions from the electron and hole surfaces to the anomalous thermal Hall conductivity is a universal result independent of the details of the model.} 
    
 {   In the case with both the electron ring-shaped Fermi surface and the hole cylinder-shaped surfaces, which is suggested by band calculations~\cite{UTe2_Yanase,UTe2_quasi2D, PhysRevB.100.134502,UTe2_Shishidou}, for a magnetic field parallel to the $a$ and $b$-axes, the-ring shaped Fermi surface dominates, while for a magnetic field parallel to the $c$-axis, both types of the Fermi surfaces give comparable contributions, leading to the substantial cancellation of the ATHE. However, we note that in this case, the cancellation is not exact even though the system is a compensated metal, because of the different dimensionality of the electron Fermi surface (three-dimensional) and the hole Femri surface (quasi-two-dimensional). We stress again that the informations of the Fermi surface are important for the determination of the pairing states from the thermal Hall measurements. }
    
    In this paper, we  consider only the intrinsic anomalous thermal Hall conductivity, which directly reflects the topology of the system and are robust against the effects of impurities. On the other hand, impurity scatterings give rise to the extrinsic anomalous thermal Hall effect.~\cite{PhysRevLett.60.2206, Yip_2016, PhysRevLett.124.157002, PhysRevResearch.2.023223} Therefore, the calculation of the extrinsic thermal Hall conductivity, which strongly depends on impurities, and its comparison with the intrinsic effect obtained in this study are an important future issue.

\begin{acknowledgment}
    The authors are grateful to D. Aoki, H. Harima, K. Izawa, K. Ishida, S. Kitagawa, T. Shibauchi for valuable discussions. This work was also supported by JST CREST Grant No. JPMJCR19T5, Japan, and the Grant-in-Aid for Scientific Research on Innovative Areas "Quantum Liquid Crystals (No. JP20H05163 and No. JP22H04480)" from JSPS of Japan,  and JSPS KAKENHI (Grant No.~JP20K03860, No.~JP20H01857, No.~JP21H01039, No. JP22H01221, and No. JP22K14005). T. M. was supported by a Japan Society for the Promotion of Science (JSPS) Fellowship for Young Scientists and by JSPS KAKENHI Grant No.~JP19J20144. M.G.Y. is supported by Multidisciplinary Research Laboratory System for Future Developments, Osaka University.
\end{acknowledgment}

\appendix

\section{Superconducting gap node}
\begin{figure}[b]
            \includegraphics[width=15cm]{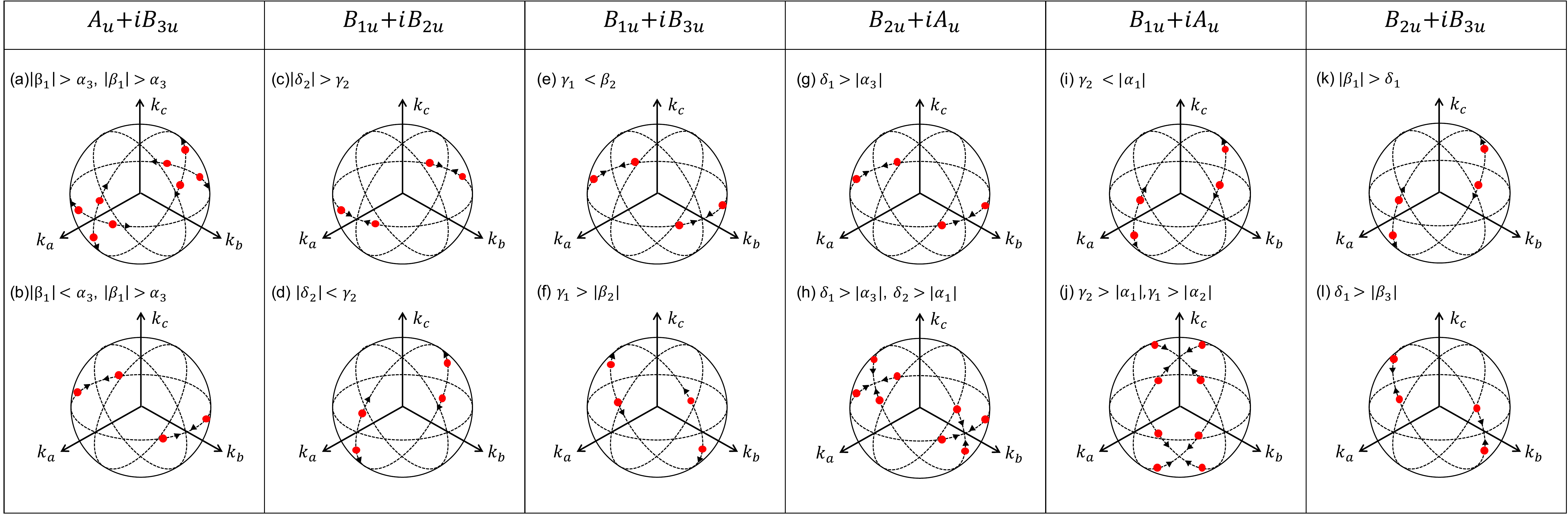}
            \centering
            \caption{Point node structures for all states considered in this paper. The red point represents the point nodes and the black triangle represents the direction of them against the increase of $r$.}
            \label{point_node_structure}
        \end{figure}

    In this section, we describe the gap node structure of UTe$_2$. The energy spectrum of the spin-triplet superconductor is given by 
    \begin{equation}
        E_{\pm}(\bm{k}) = \sqrt{\epsilon^2(\bm{k})) + |\bm{d}(\bm{k})|^2 \pm |\bm{d}(\bm{k}) \times \bm{d}^*(\bm{k})|},
        \label{eqs6}
    \end{equation}
    and the condition for the appearance of gap nodes is given by
    \begin{equation}
        |\bm{d}(\bm{k})|^2 -|\bm{d}(\bm{k}) \times \bm{d}^*(\bm{k})| = 0\\
        \Rightarrow d_a^2(\bm{k}) + d_b^2(\bm{k}) + d_c^2(\bm{k}) = 0.
        \label{eqs7}
    \end{equation}
    From this equation and equations Eq.(\ref{eq3})-(\ref{eq5}), the position of point nodes for each $d$-vector are as follows. The schematic of the node structure for each case is shown in Fig.\ref{point_node_structure}.
    \subsection*{The $A_u + iB_{3u}$ state}
        \begin{equation}
            \begin{split}
                (\mathrm{i}) \ \bm{k}_{(\mathrm{i})} = \left(k_{a(\mathrm{i})}, 0, \pm \frac{\alpha_1}{\sqrt{|\beta_1|^2 - \alpha_3^2}}k_{a(\mathrm{i})} \right)\\
                (\mathrm{ii}) \ \bm{k}_{(\mathrm{ii})} = \left(k_{a(\mathrm{ii})},\pm \frac{\alpha_1}{\sqrt{|\beta_2|^2 - \alpha_2^2}}k_{a(\mathrm{ii})}, 0 \right)
            \end{split}
            \label{eqs8}
        \end{equation}
        Here, we take $\alpha_i\ (i=1,2,3)$ as real and $\beta_i\ (i=1,2)$ as pure imaginary. The condition (i) leads to 4 point nodes on $k_a k_c$-plane provided that $|\beta_1| > \alpha_3$. Here, $k_{a(\mathrm{i})}$ is determined by crossing points of the line $k_b = 0,\ k_c = \pm\frac{\alpha_1}{\sqrt{|\beta_1|^2 - \alpha_3^2}}k_a$ and the Fermi surface. The condition (ii) leads to 4 point nodes on $k_a k_b$-plane provided that $|\beta_2| > \alpha_2$. Here, $k_{a(\mathrm{ii})}$ is determined by crossing points of the line $k_c = 0,\ k_b = \pm\frac{\alpha_1}{\sqrt{|\beta_2|^2 - \alpha_2^2}}k_a$ and the Fermi surface. These two conditions is determined by the magnitude correlation of coefficients $|\beta_1| > |\alpha_3|$ and $|\beta_2| > |\alpha_2|$ independently. Therefore, when these two inequality are satisfied simultaneously, 8 point nodes can appear.
    \subsection*{The $B_{1u} + iB_{2u}$ state}
        \begin{equation}
            \begin{split}
                (\mathrm{ii}) \ \bm{k}_{(\mathrm{iii})} = \left(\pm \frac{\gamma_1}{\sqrt{|\delta_2|^2 - \gamma_2^2}}k_{b(\mathrm{iii})},k_{b(\mathrm{iii})},0 \right)\\
                (\mathrm{iv}) \ \bm{k}_{(\mathrm{iv})} = \left(\pm \frac{|\delta_1|}{\sqrt{\gamma_2^2 - |\delta_2|^2}}k_{c(\mathrm{iv})}, 0, k_{c(\mathrm{iv})} \right)
            \end{split}
            \label{eqs9}
        \end{equation}
        Here, we take $\gamma_i\ (i=1,2)$ as real and $\delta_i\ (i=1,2)$ as pure imaginary. The condition (iii) leads to 4 point nodes on $k_a k_b$-plane provided that $|\delta_2| > \gamma_2$. Here, $k_{b(\mathrm{iii})}$ is determined by crossing points of the line $k_a = \pm \frac{\gamma_1}{\sqrt{|\delta_2|^2 - \gamma_2^2}}k_b,\ kc = 0$ and the Fermi surface. The condition (iv) leads to 4 point nodes on $k_a k_c$-plane provided that $\gamma_2 > |\delta_2|$. Here, $k_{c(\mathrm{iii})}$ is determined by crossing points of the line $k_a = \pm \frac{|\delta_1|}{\sqrt{\gamma_2^2 - |\delta_2|^2}}k_c,\ k_b = 0$ and the Fermi surface. Unlike the case of the $A_u + iB_{3u}$ state, these two conditions are only determined by $\gamma_2$ and $\delta_2$, so there are always four point nodes.
    \subsection*{The $B_{1u} + iB_{3u}$ state}
        \begin{equation}
            \begin{split}
                (\mathrm{v}) \ \bm{k}_{(\mathrm{v})} = \left(0, \pm \frac{|\beta_1|}{\sqrt{\gamma_1^2 - |\beta_2|^2}}k_{c(\mathrm{v})}, k_{c(\mathrm{v})}, \right)\\
                (\mathrm{vi}) \ \bm{k}_{(\mathrm{vi})} = \left(k_{a(\mathrm{vi})},\pm \frac{\gamma_2}{\sqrt{|\beta_2|^2 - \gamma_1^2}}k_{a(\mathrm{vi})}, 0 \right)
            \end{split}
            \label{eqs10}
        \end{equation}
        Here, we take $\gamma_i\ (i=1,2)$ as real and $\beta_i\ (i=1,2)$ as pure imaginary. The condition (v) leads to 4 point nodes on $k_b k_c$-plane provided that $\gamma_1 > |\beta_2|$. Here, $k_{c(\mathrm{v})}$ is determined by crossing points of the line $k_a = 0,\ k_b = \pm \frac{|\beta_1|}{\sqrt{\gamma_1^2 - |\beta_2|^2}}k_c$ and the Fermi surface. The condition (vi) leads to 4 point nodes on $k_a k_b$-plane provided that $|\beta_2| > \gamma_1$. Here, $k_{a(\mathrm{vi})}$ is determined by crossing points of the line $k_c = 0,\ k_b = \pm \frac{\gamma_2}{\sqrt{|\beta_2|^2 - \gamma_1^2}}k_a$ and the Fermi surface. Unlike the case of the $A_u + iB_{3u}$ state, these two conditions are only determined by $\gamma_1$ and $\beta_2$, so there are always four point nodes.
    \subsection*{The $B_{2u}+iA_u$ state}
    \begin{equation}
        \begin{split}
            (\mathrm{vii}) \ \bm{k}_{(\mathrm{vii})} = \left(0, k_{b(\mathrm{vii})}, \pm \frac{|\alpha_2|}{\sqrt{\delta_2^2 - |\alpha_1|^2}}k_{b(\mathrm{vii})} \right)\\
            (\mathrm{viii}) \ \bm{k}_{(\mathrm{viii})} = \left(\pm \frac{|\alpha_2|}{\sqrt{\delta_1^2 - |\alpha_3|^2}}k_{b(\mathrm{viii})}, k_{b(\mathrm{viii})}, 0 \right)
        \end{split}
        \label{eqs11}
    \end{equation}
        Here, we take $\delta_i\ (i=1,2)$ as real and $\alpha_i\ (i=1,2,3)$ as pure imaginary. The condition (vii) leads to 4 point nodes on $k_b k_c$-plane provided that $\delta_2 > |\alpha_2|$. Here, $k_{b(\mathrm{vii})}$ is determined by crossing points of the line $k_a = 0,\ k_c = \pm \frac{|\alpha_2|}{\sqrt{\delta_2^2 - |\alpha_1|^2}}k_b$ and the Fermi surface. The condition (viii) leads to 4 point nodes on $k_a k_b$-plane provided that $\delta_1 > |\alpha_3|$. Here, $k_{b(\mathrm{viii})}$ is determined by crossing points of the line $k_c = 0,\ k_a = \pm \frac{|\alpha_2|}{\sqrt{\delta_1^2 - |\alpha_3|^2}}k_b$ and the Fermi surface. In this case, when two inequality, $\delta_2 > |\alpha_2|$ and $\delta_1 > |\alpha_3|$, are satisfied simultaneously, 8 point nodes can appear.
    \subsection*{The $B_{1u}+iA_u$ state}
    \begin{equation}
        \begin{split}
            (\mathrm{ix}) \ \bm{k}_{(\mathrm{ix})} = \left(0, \pm \frac{|\alpha_3|}{\sqrt{\gamma_1^2 - |\alpha_2|^2}}k_{c(\mathrm{ix})}, k_{c(\mathrm{ix})} \right)\\
            (\mathrm{x}) \ \bm{k}_{(\mathrm{x})} = \left(\pm \frac{|\alpha_3|}{\sqrt{\gamma_2^2 - |\alpha_1|^2}}k_{c(\mathrm{x})}, 0, k_{c(\mathrm{x})} \right)
        \end{split}
        \label{eqs12}
    \end{equation}
            Here, we take $\gamma_i\ (i=1,2)$ as real and $\alpha_i\ (i=1,2,3)$ as pure imaginary. The condition (ix) leads to 4 point nodes on $k_b k_c$-plane provided that $\gamma_1 > |\alpha_2|$. Here, $k_{b(\mathrm{x})}$ is determined by crossing points of the line $k_a = 0,\ k_b = \pm \frac{|\alpha_3|}{\sqrt{\gamma_1^2 - |\alpha_2|^2}}k_c$ and the Fermi surface. The condition (x) leads to 4 point nodes on $k_a k_c$-plane provided that $\gamma_2 > |\alpha_1|$. Here, $k_{b(\mathrm{ix})}$ is determined by crossing points of the line $k_b = 0,\ k_a = \pm \frac{|\alpha_3|}{\sqrt{\gamma_2^2 - |\alpha_1|^2}}k_c$ and the Fermi surface. In this case, when two inequality, $\gamma_2 > |\alpha_1|$ and $\gamma_1 > |\alpha_2|$, are satisfied simultaneously, 8 point nodes can appear.
    \subsection*{The $B_{2u} + iB_{3u}$ state}
        \begin{equation}
            \begin{split}
                (\mathrm{xi}) \ \bm{k}_{(\mathrm{xi})} = \left(0, k_{b(\mathrm{xi})}, \pm \frac{|\beta_2|}{\sqrt{\delta_1^2 - |\beta_1|^2}}k_{b(\mathrm{xi})} \right)\\
                (\mathrm{xii}) \ \bm{k}_{(\mathrm{xii})} = \left(k_{a(\mathrm{xii})}, 0, \pm \frac{\delta_2}{\sqrt{|\beta_1|^2 - \delta_1^2}}k_{a(\mathrm{xii})} \right)
            \end{split}
            \label{eqs13}
        \end{equation}
        Here, we take $\delta_i\ (i=1,2)$ as real and $\beta_i\ (i=1,2)$ as pure imaginary. The condition (xi) leads to 4 point nodes on $k_b k_c$-plane provided that $\delta_1 > |\beta_2|$. Here, $k_{b(\mathrm{xi})}$ is determined by crossing points of the line $k_a = 0,\ k_b = \pm \frac{|\beta_2|}{\sqrt{\delta_1^2 - |\beta_1|^2}}k_c$ and the Fermi surface. The condition (xii) leads to 4 point nodes on $k_a k_c$-plane provided that $|\beta_1| > \delta_1$. Here, $k_{a(\mathrm{xii})}$ is determined by crossing points of the line $k_b = 0,\ k_c = \pm \frac{\delta_2}{\sqrt{|\beta_1|^2 - \gamma_1^2}}k_a$ and the Fermi surface. Unlike the case of the $B_{2u} + iB_{3u}$ state, these two conditions are only determined by $\beta_1$ and $\delta_2$, so there are always four point nodes.

\bibliography{UTe2_IATHE_ref}
\bibliographystyle{jpsj}

\end{document}